\documentclass[%
 aip,
 amsmath,amssymb,
preprint,
]{revtex4-1}

\usepackage{graphicx}
\usepackage{dcolumn}
\usepackage{bm}

\usepackage[utf8]{inputenc}
\usepackage[T1]{fontenc}
\usepackage{mathptmx}
\usepackage{etoolbox}
\usepackage{color}
\makeatletter
\def\@email#1#2{%
 \endgroup
 \patchcmd{\titleblock@produce}
  {\frontmatter@RRAPformat}
  {\frontmatter@RRAPformat{\produce@RRAP{*#1\href{mailto:#2}{#2}}}\frontmatter@RRAPformat}
  {}{}
}%
\makeatother
\begin{document}

\preprint{}

\title[]{Effect of a fixed downstream cylinder on the flow-induced vibration of an elastically-supported primary cylinder}

\author{Junlei Wang}

\author{Shenfang Li}
\affiliation{School of Mechanical and Power Engineering, Zhengzhou University, Zhengzhou 450000, China
}

\author{Daniil Yurchenko}
\affiliation{Institute of Sound and Vibration Research, University of Southampton, Southampton, SO17, United Kingdom
}

\author{Hongjun Zhu}
\affiliation{State Key Laboratory of Oil and Gas Reservoir Geology and Exploitation, Southwest Petroleum University, Chengdu 610500, China}

\author{Chandan Bose}
\email{Chandan Bose - c.bose@bham.ac.uk}
\affiliation{Aerospace Engineering, School of Metallurgy and Materials, The University of Birmingham, Birmingham, B15 2TT, United Kingdom}%


\begin{abstract}
This paper numerically investigates the influence of a fixed downstream control cylinder on the flow-induced vibration of an elastically-supported primary cylinder. These two cylinders are situated in a tandem arrangement with small dimensionless centre-to-centre spacing ($L/D$, $L$ is the intermediate spacing, and $D$ is the cylinder diameter). The present two-dimensional (2D) simulations are carried out in the low Reynolds number ($Re$) regime. The primary focus of this study is to reveal the underlying flow physics behind the transition from vortex-induced vibration to galloping in the response of the primary cylinder due to the presence of another fixed downstream cylinder. Two distinct flow field regimes, namely steady flow and alternate attachment regimes, are observed for different $L/D$ and Re values. Depending on the evolution of the near-field flow structures, four different wake patterns - `2S', `2P', `2C',  and `aperiodic' - are observed. The corresponding vibration response of the upstream cylinder is characterized as interference galloping and extended vortex-induced vibration. As the $L/D$ ratio increases, the lift enhancement due to flow-induced vibration is seen to be weakened. The detailed correlation between the force generation and the near-wake interactions is investigated. The present findings will augment the understanding of vibration reduction or flow-induced energy harvesting of tandem cylindrical structures.
\end{abstract}

\maketitle

\section{\label{sec:intro}Introduction}

Slender cylindrical structures are commonly used in various engineering applications, such as cables of suspension bridges, pipelines of offshore wells, electric lines, masts, and chimneys, among others. Flow-induced vibration (FIV) of such bluff-body-shaped engineering structures is ubiquitous and has received extensive attention in recent decades \citep{chen2021flow, hu2019performance,zafar2018low}.
The separation and rolling up of shear layers give way to vortex shedding behind cylinders, which in turn causes the pulsating forces on them, thereby inducing FIV \citep{zhang2022spacing, zhang2022vortex}. When the vortex shedding frequency is close to the structural natural frequency, synchronization takes place, giving rise to high-amplitude vibrations in the lock-in regime. This phenomenon is called vortex-induced vibration (VIV) \citep{williamson2004vortex}. Galloping is another kind of aeroelastic instability, where the resulting self-sustained response amplitude increases with the increasing reduced velocity $U_r$. Galloping usually occurs on structures with sharp edges as they are prone to aeroelastic instabilities. Interestingly, adding additional surface protrusions to the surface of a circular cylinder can also give rise to galloping \citep{wang2020hybrid,wang2021enhancing}. Further, in tandem arrangement with a considerably small spacing ratio, the downstream cylinder (DC) can induce interference galloping in the response of the upstream cylinder (UC) \citep{assi2006experimental}.

In the presence of multiple cylinders, the FIV characteristics get significantly altered compared to a single cylinder due to their mutual interference. An elastically mounted cylinder is likely to exhibit wake galloping due to the strong disturbance coming from the wake of another cylinder placed in the near-field \citep{khan2022flow}. Previous studies have investigated amplitude and frequency characteristics of the response, different flow field regimes, vortex shedding patterns, and aerodynamic force generation of the wake galloping phenomenon \citep{assi2010suppression,assi2010wake, bakhtiari2020effect, tang2022numerical}. \citet{ping2022vortex} reported that the lower response branch of rigidly connected tandem cylinders is quasi-periodic, and it can be divided into two different sub-branches due to the frequency detuning. \citet{zhu2019wake} numerically studied the effect of an UC and spacing ratio on the motion of a DC. The authors reported that the wake pattern changed significantly with the varying spacing ratios and reduced velocities. The downstream circular cylinder exhibited an aperiodic response due to the unstable vortex evolution generated by the rectangular and triangular cylinders. \citet{qin2017two} investigated the FIV of different-sized tandem cylinders, in which the small UC was fixed, and the large DC was elastically supported, showing that both vortex shedding and gap shear layer switching could lead to a rapid increase in vibration during the initial transition. \citet{hu2020flow} reported the vibration response of the circular cylinder in the wake of an UC; three vibration responses of the DC, namely pure vortex resonance, separated pure vortex resonance and wake-induced galloping, and combined pure vortex resonance and wake-induced galloping, were observed. Unlike the vortex shedding vibration mechanism for a single cylinder, there are other vibration excitation mechanisms in tandem cylinders. \citet{borazjani2009vortex} conducted a study on elastically supported cylinders arranged in series, and they highlighted the significant role of gap flow in initiating and sustaining vibrations in the UC. The gap flow generated substantial oscillatory forces, resulting in the cylinder maintaining large amplitude and low-frequency oscillations. 

Adding more cylinders increases complexity in the flow field, including alterations in the shear layer within the gaps and the merging of vortices in the flow field. Consequently, the FIV behaviour of the cylinders also varies. \citet{zhu2021wake} studied the fluid force characteristics of three tandem cylinders and pointed out that the fluid forces of the DC are reduced during the transition from continuous reattachment-alternate reattachment to quasi-co-shedding. \citet{tu2022study} numerically studied the FIV characteristics of three cylinders in series by varying $Re$, $U_r$ and the shear ratio and showed that the FIV response of the middle and DCs is more sensitive to the parameters mentioned above. When the spacing ratio is 5.5, the interference of the DC on the UC is reduced, and the vibration of the UC closely resembles that of a single cylinder. \citet{duong2022low} numerically investigated the wake flow regime of three elliptical cylinders in series and showed that two kinds of vortex shedding frequencies appear in the middle and DCs while the flow regime is quasi-co-shedding (the vortices behind the DC primarily occur due to the vortex shedding from the UC) and co-shedding (vortex shedding takes place simultaneously behind three cylinders). \citet{hosseini2022flow} investigated the FIV of six elastically supported cylinders with equal spacing. At a spacing ratio of 3, the flow becomes disordered as the reduced velocity increases. The vibration of the primary cylinder closely resembles that of a single cylinder, while the vibration of the other cylinders exhibits random variations with reduced velocity.

Previous studies have primarily concentrated on assessing the impact of a series of elastically supported cylinders or the wake-induced vibration of the DC. The influence of DC on the FIV of UC has received the least attention. However, it has been shown by a few studies that a DC can give rise to nonlinear and divergent self-excited vibration similar to galloping in the UC cylinder response \citep{sumner2010two}. \citet{zhu2019wake} investigated the effect of the C-shaped DC on the vibration feature of the UC. The results show that at $Re = 100$ and the spacing equals to $1.5D$, the UC exhibits the extended body feature; the lift coefficient and amplitude of the UC are reduced by 85.5\%, and 94.5\% due to the interference of the C-shaped cylinder. \citet{zhang2017improving} studied the energy harvesting performance of FIV of tandem cylinders with a gap ratio between 0.3 to 3, in which the DC is fixed and the UC is elastically supported. The experimental results show that the resonance region of the UC is broadened by setting the interference cylinder. In addition, \citet{zhang2019experimental} experimentally investigated the effect of the spacing and the shapes of the fixed DC on the FIV of the UC. The gap ratio was from 0.15 to 4.5. It is pointed out that when the DC is a square, the synchronous region is increased by 380\% compared with that of a single cylinder. However, these studies did not conduct an in-depth analysis of the underlying reason why the presence of the fixed downstream interference cylinder could enhance the vibration of the UC at a low spacing ratio, which will be taken up in this paper. 

Summarizing the previous works, many researchers have concentrated on the FIV of the DC in the wake flow. Although a few studies have demonstrated the effect of fixed DC on the UC, the underlying flow mechanisms behind the transition from VIV to galloping of the UC are yet to be clearly understood. Therefore, it is necessary to conduct a systematic, in-depth study to understand the interference mechanism of the fixed DC, especially when the intermediate spacing is small. This is taken up in the present study. Note that the present 2D simulations are carried out in the low $Re$ regime to avoid the effect of three-dimensionality. Due to the relatively lower computational cost of 2D simulations, many previous studies also have chosen to study the FIV of bluff-body systems in the low $Re$ regime to capture the essential flow physics and have shown that low $Re$ simulations are of great significance in unravelling the underlying fundamental vortex dynamics behind the FIV of bluff-body shaped structures. We aim to investigate the following key research questions: (1) what are the wake characteristics for a range of different $Re$ and spacing ratios in the tandem circular cylinder? (2) what are the vibration characteristics and the FIV transition mechanism of the UC for different $Re$ and $L/D$ values? (3) what is the role of the wake interference in the intermediate gap between the tandem cylinders in changing the hydrodynamic coefficient of the UC? A series of systematic 2D laminar simulations are carried out in this study to address these questions. 

The remainder of this paper is organized as follows. Section 2 comprises of the details of the computational methodology, including the governing equations, the simulation set-up, and the verification and validation of the fluid-structure interaction solver. The flow-field characteristics of the tandem cylinder system are depicted in Section 3, while Sections 4 and 5 present the flow-induced vibration and force generation characteristics of the primary cylinder, respectively. Finally, the salient outcomes of this study are summarized in Section 6.

\section{Computational Methodology}

\subsection{Problem Definition}

\begin{figure}[htbp]
    \begin{center}
    \includegraphics[width=0.6\textwidth]{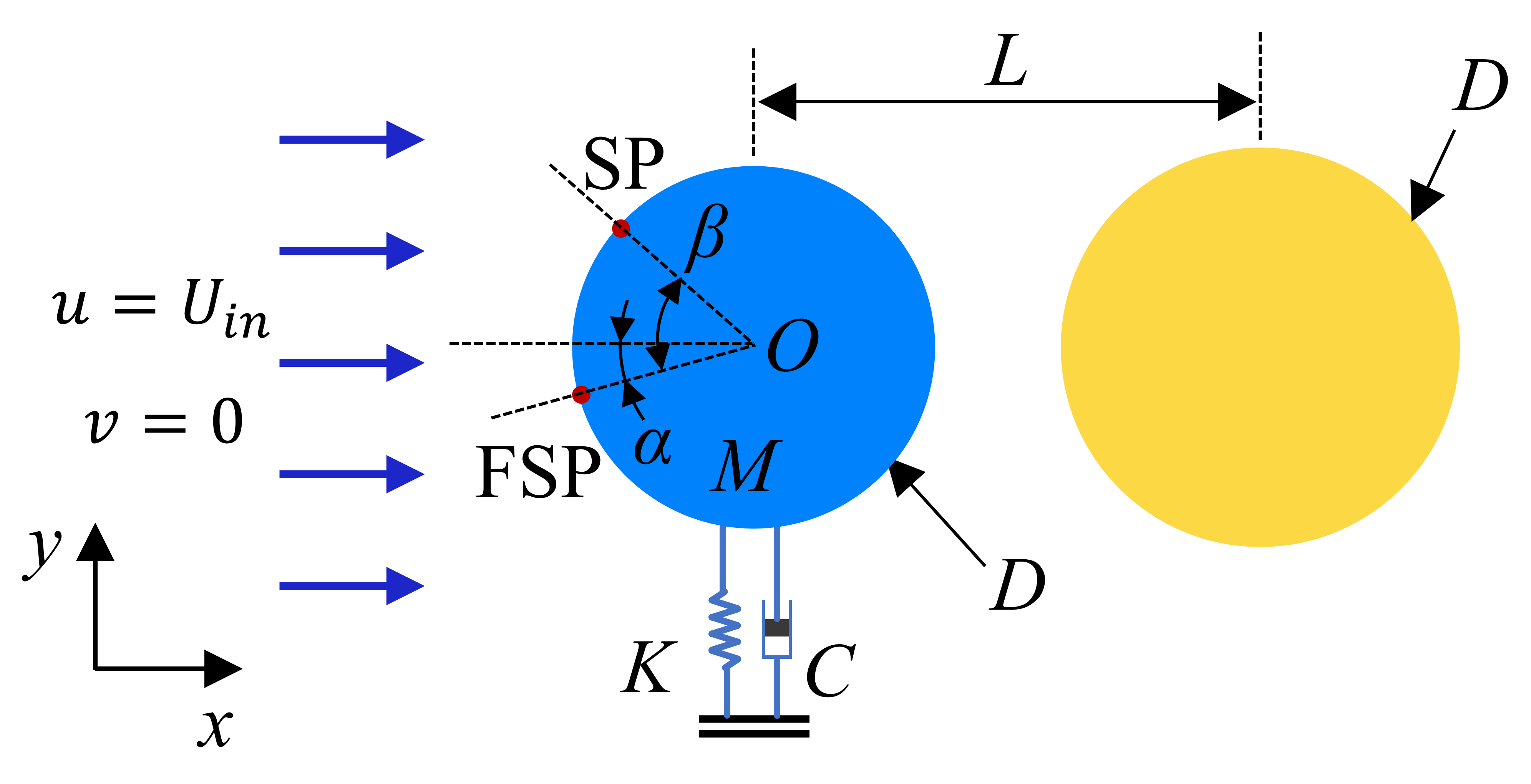} 
    \caption{Schematic of the tandem cylinder system considered in the present study.}
    \label{fig:fig1}
    \end{center}
\end{figure}

The FIV of an upstream elastically supported cylinder with a downstream fixed control cylinder in the low Reynolds number regime is of present interest; see the problem schematic in Fig.~\ref{fig:fig1}. The single-degree-of-freedom elastically supported circular cylinder is located upstream and can move freely along the $y$ direction. The fixed control circular cylinder is located downstream. The initial state of the two cylinders is on the same horizontal line. The nondimensional distance between the centers of the two cylinders, $L/D$, is systematically varied from $1.25$ to $3$, where $D$ is the diameter of the circular cylinder. 

\hspace{0.5cm} To facilitate the analysis in the later sections, Fig.~\ref{fig:fig1} shows the front stagnation point of the cylinder marked as FSP and the separation point of the boundary layer marked as SP. The angle $\alpha$ is defined as the angle between the front stagnation point and the center horizontal line of the cylinder. The angle is positive when the front stagnation point is above the center horizontal line of the cylinder, and negative when it is below. The angle $\beta$ is the angle between the separation point of the boundary layer and the front stagnation point.

\subsection{Governing equations}

The viscous Newtonian flow is governed by the continuity and unsteady incompressible Navier-Stokes equations as given in the dimensionless form in Eqs. 1 and 2.

\begin{equation} \label{eq:continuity}
   \frac{\partial u_i}{\partial x_i} = 0,
\end{equation}

\begin{equation} \label{eq:NS}
 \frac{\partial u_i}{\partial t} +  u_j\frac{\partial u_i}{\partial x_j} =  -\frac{\partial p}{\partial x_i} + \frac{1}{Re} \frac{\partial^2 u_i}{\partial x_i \partial x_j},
\end{equation}

\noindent where $t$ ($= \hat{t} U_{in}/D$) denotes the dimensionless time, $u_i$ ($= \hat{u}_i/U_{in}$) are the dimensionless velocity components, $x_i$ ($=\hat{x}_i/D$) are the dimensionless coordinates, $p$ ($=\hat{p}/\rho U_{in}^2$) is the dimensionless pressure, and $Re$ ($ = U_{in}D/{\nu}$) is the Reynolds number. Here, $D$ is the diameter of the cylinder, $U_{in}$ is the free-stream velocity, $\hat{t}$ is the dimensional time, $\hat{x_i}$ are the dimensional coordinates, $\hat{u_i}$, $\hat{p}$, $\rho$, $\nu$ are the dimensional velocity components, pressure, density, and kinematic viscosity of the fluid, respectively. The finite volume method is used to solve the incompressible flow governing equations. The convection terms are discretized in a second-order upwind scheme, and the pressure and velocity coupling are solved by the SIMPLE algorithm. It is known that for a single cylinder subjected to uniform flow, the vortex shedding occurs when $Re > 40$ and the wake becomes three-dimensional when $Re \sim 200$\citep{williamson1988existence}. Hence, the present study is focused on the low $Re$ regime; $Re$ ranging from 50 to 170.

\hspace{0.5cm} The structural system consists of a fixed DC and an elastically supported UC with a single degree of freedom (sdof) along the cross-flow direction. A typical mass-spring-damper dynamic model describes the motion characteristics of the elastically supported circular cylinder. The motion of the UC in the `y' direction is governed by a second-order linear differential equation describing the dynamics of a sdof system as follows:

\begin{equation}
    M\ddot{\xi}+C\dot{\xi}+K\xi=F_y,
\end{equation}

\noindent where $M$ represents the structural mass, $K$ represents the system stiffness, and $\sqrt{\frac{K}{M}}=\omega_n = 2 \pi f_n$, where $f_n$ is the structural natural frequency. $\xi$ is the displacement of the UC, and $F_y$ denotes the lift force along the `y' direction acting on the cylinder. Equation 3 can be non-dimensionalised using the diameter $D$ and the free stream velocity $U_{in}$ as the length and velocity scales, respectively, as follows:

\begin{equation}
    \ddot{Y}+\frac{4 \pi \zeta}{U_r} \dot{Y}+ \frac{4 \pi^2}{U_r^2} Y = \frac{C_L}{2m^*},
\end{equation}

\noindent where nondimensional displacement $Y = \xi/D$, damping ratio $\zeta = C/2\sqrt{MK}$, reduced velocity $U_r = \frac{U_{in}}{f_n D} = \frac{Re \nu}{f_n D^2}$, and mass ratio $m^* = M/\rho D^2$. Here, $C$ is the structural damping coefficient, $K$ is the structural stiffness), $M$ is the structural mass of the cylinder per unit length, and $f_n$ is the structural natural frequency. In this study, $U_r$ is proportional to $Re$ as the natural frequency of the structure $f_n$, kinematic viscosity of the fluid $\nu$, and the cylinder diameter $D$ are kept constant throughout the study. The $m^*$ and $\zeta$ of the oscillating cylinder system are considered to be 1.113 and 0.01, respectively, following \citet{ZHOU1999165}. The structural governing equation is numerically solved using the fourth-order Runge-Kutta method.

The two-way coupled fluid-structure interaction (FSI) simulations are carried out using the CFD software Ansys Fluent based on the fully--implicit algorithm, where the structural equations are incorporated through user-defined functions. To calculate the two-way coupled response of the UC, the N-S equations are coupled with the structural governing equations in a staggered manner.  The second-order accurate spatial and temporal discretization schemes are used in this study with the absolute and relative tolerance criteria kept at $10^{-6}$. 

\subsection{Verification and Validation}

\begin{figure}[htbp]
    \begin{center}
    \includegraphics[width=0.9\textwidth]{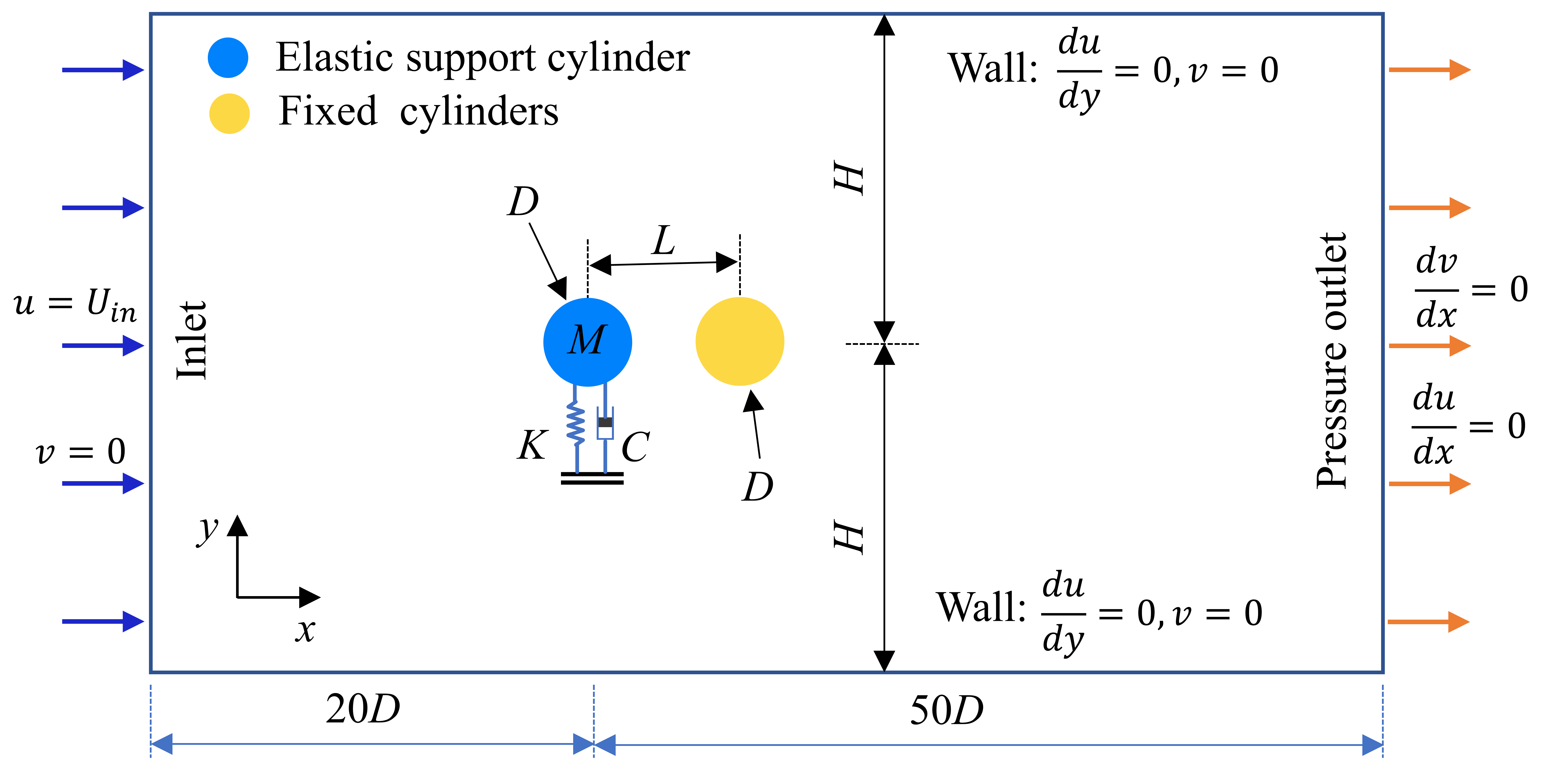} 
    \caption{Schematic of the present computational domain, and imposed boundary conditions.}
    \label{fig:fig2}
    \end{center}
\end{figure}

The entire computational domain is rectangular with a size of $70D \times 2H$, where $D$ is the diameter of the circular cylinder and $H$ is half of the width of the flow field. In the flow direction, the distance between the center point of the elastically supported cylinder and the inlet is $20D$, while the distance is $50D$ from the center of the elastically supported cylinder to the outlet. The boundary conditions are shown in Fig.~\ref{fig:fig2}. The boundary settings of the computational domain include the velocity inlet with $u = U_{in}$ and $v = 0$, the pressure outlet with $du/dx=0$ and $dv/dx=0$, and the symmetric upper and lower walls with $du/dy=0, v= 0$. 

\begin{figure}[htbp]
    \begin{center}
    \includegraphics[width=0.7\textwidth]{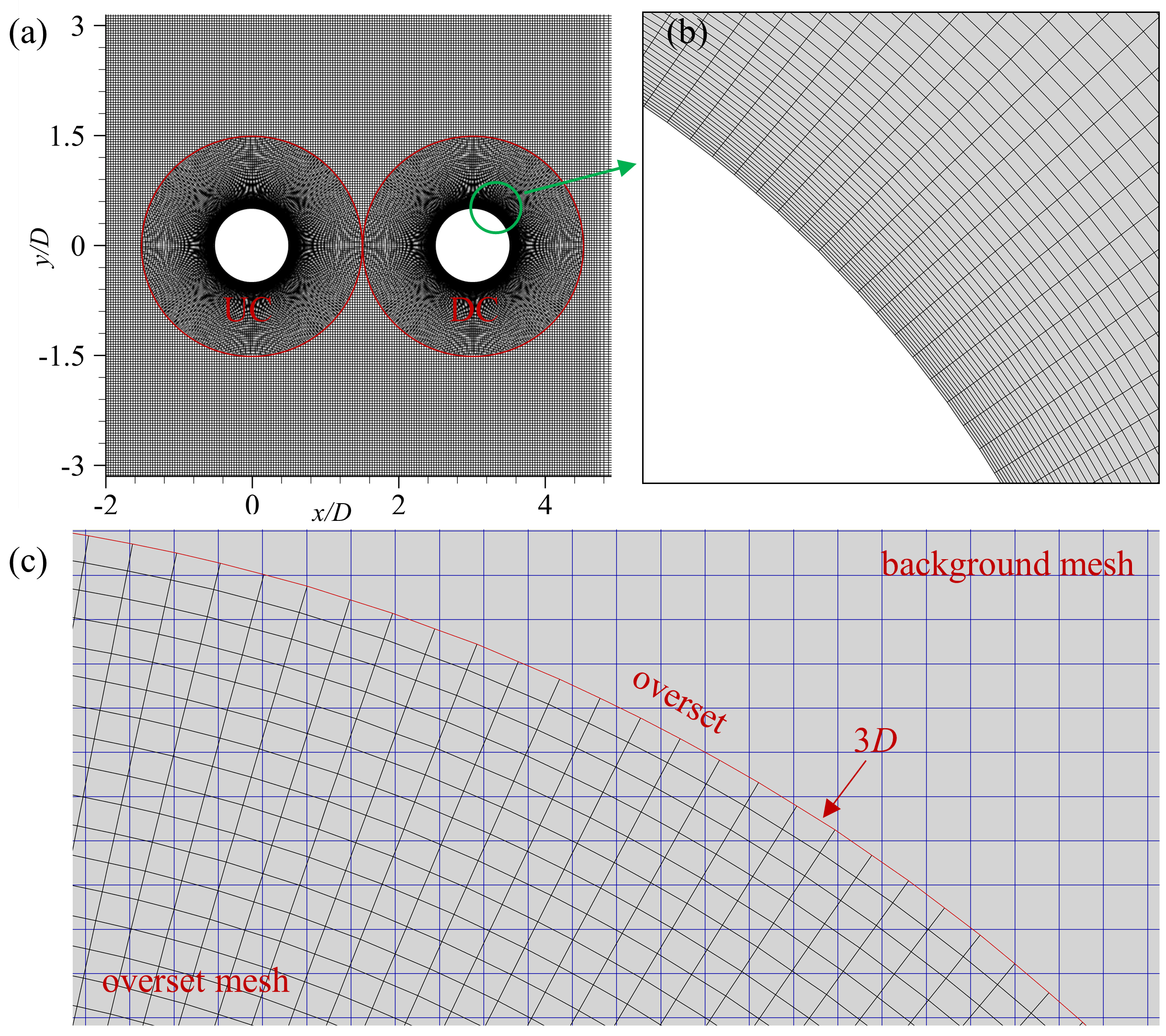} 
    \caption{Computational mesh: (a) the mesh around the tandem cylinder at its initial position; (b) zoomed inset of the mesh near the cylinder; (c) the overset meshing strategy.}
    \label{fig:fig3}
    \end{center}
\end{figure}

\hspace{0.5cm} To avoid the numerical instability caused by the excessive mesh deformation due to the large displacement of the upstream elastically supported cylinder, the overset meshing strategy is used in the present study. The computational mesh is shown in Fig.~\ref{fig:fig3}, which consists of a background mesh and two overset meshes. The partial mesh of the tandem cylinder is shown in Fig.~\ref{fig:fig3}(a), and Fig.~\ref{fig:fig3}(b) shows an enlarged view of the cylindrical boundary grid. The boundary of the overset mesh domain is $3D$, as seen in Fig.~\ref{fig:fig3}(c). To accurately capture the flow-field structures, the perimeter of the cylinder is discretized at equal intervals, and the thickness of the first layer grid around the cylinder structure and the mesh growth ratio in the radial direction are $0.001D$ and $1.02D$, respectively.
Additionally, the same grid density is used for both overset grid domains.

\hspace{0.5cm} The effect of domain size on the results is checked by varying the cross-flow height $H$ ($20D$, $25D$, and $28D$) at $L/D = 3$ and $Re = 100$. The same foreground mesh for the cylinder walls and the same grid size growth rate are used in these three cases. The time step in the domain-independence study is set to 0.008, meeting the requirement of the maximum Courant number to be less than unity to ensure the stability of the solution scheme. The simulation results are shown in Table 1, the maximum difference is less than $1\%$. To reduce the computational cost, the $H$ value of $20D$ is considered by comparing the results of $Y_{max}$, ${C_D}_{mean}$, and ${C_L}_{rms}$ for these three cases.

\begin{table}
    \centering
    \caption{Results comparison under different heights of the computational domain for $L/D=3, Re=100$.}
    \begin{tabular}{c c c c c c c}
       $H/D$  & Background elements & Overset elements & $\Delta t$ & $Y_{max}$ & ${C_D}_{mean}$ & ${C_L}_{rms}$\\
       \hline 
       \hline
         $20$ & 80896 & 49128 & 0.008 &0.595 [-] & 1.7309 [-] & 0.1796 [-] \\
         $25$ & 98784 & 49128 & 0.008 &0.596 [0.17\%] & 1.7353 [0.25\%] & 0.1799 [0.17\%]\\
         $28$ & 102144 & 49128 & 0.008 &0.596 [0.00\%] & 1.7418 [0.37\%] & 0.1799 [0.00\%]\\
    \end{tabular}
    \label{tab:tab1}
\end{table}

\hspace{0.5cm} Next, the mesh size and time-step independence studies are carried out, keeping the computational domain fixed at $70D \times 40D$ at $L/D = 3$ and $Re = 100$. Case 1, Case 2, and Case 3 are used to check for the grid convergence; and Case 2 and Case 4, with Case 2 being the reference case, are used for non-dimensional time step study. The comparison of vibration response results is presented in Table 2. The number of mesh elements of the three groups is gradually increased. Comparing Case 2 and Case 1, the differences in the maximum dimensionless amplitude $Y_{max}$, the mean value of the drag coefficient ${C_D}_{mean}$, and the root mean square of the lift coefficient ${C_L}_{rms}$ are found to be 0.17\%, 0.04\%, and 0.45\%, respectively. When the number of elements is further increased to achieve Case 3, the differences are reduced to 0\%, 0.08\%, and 0.11\%, respectively. In the time-step independence results, the difference is less than 0.2\%. Based on these results, the Case 2 mesh with a non-dimensional time step of 0.003 is selected for further simulations in this paper. 

\begin{table}
    \centering
        \caption{Results of mesh independence test for flow around two tandem circular cylinders with $L/D=3$, and $Re=100$ (considering $H/D = 20$).}
    \begin{tabular}{c c c c c c c}
       Case  & Background elements & Overset elements & $\Delta t$ & $Y_{max}$ & ${C_D}_{mean}$ & ${C_L}_{rms}$\\
       \hline 
         Case 1  & 80896 & 42008 & 0.003 & 0.594 [-] & 1.7492 [-] & 0.1784 [-]\\
         Case 2 & 92736 & 49128 & 0.003 & 0.593 [0.17\%] & 1.7499 [0.04\%] & 0.1792 [0.45\%] \\
         Case 3  & 106029 & 56248 & 0.003 & 0.593 [0.00\%] & 1.7513 [0.08\%] & 0.1794 [0.11\%]\\
         Case 4 & 92736 & 49128 & 0.001 & 0.592 [0.17\%] & 1.7516 [0.10\%] & 0.1791 [0.06\%]\\
    \end{tabular}
    \label{tab:tab2}
\end{table}

\begin{figure}[htbp]
    \begin{center}
    \includegraphics[width=0.6\textwidth]{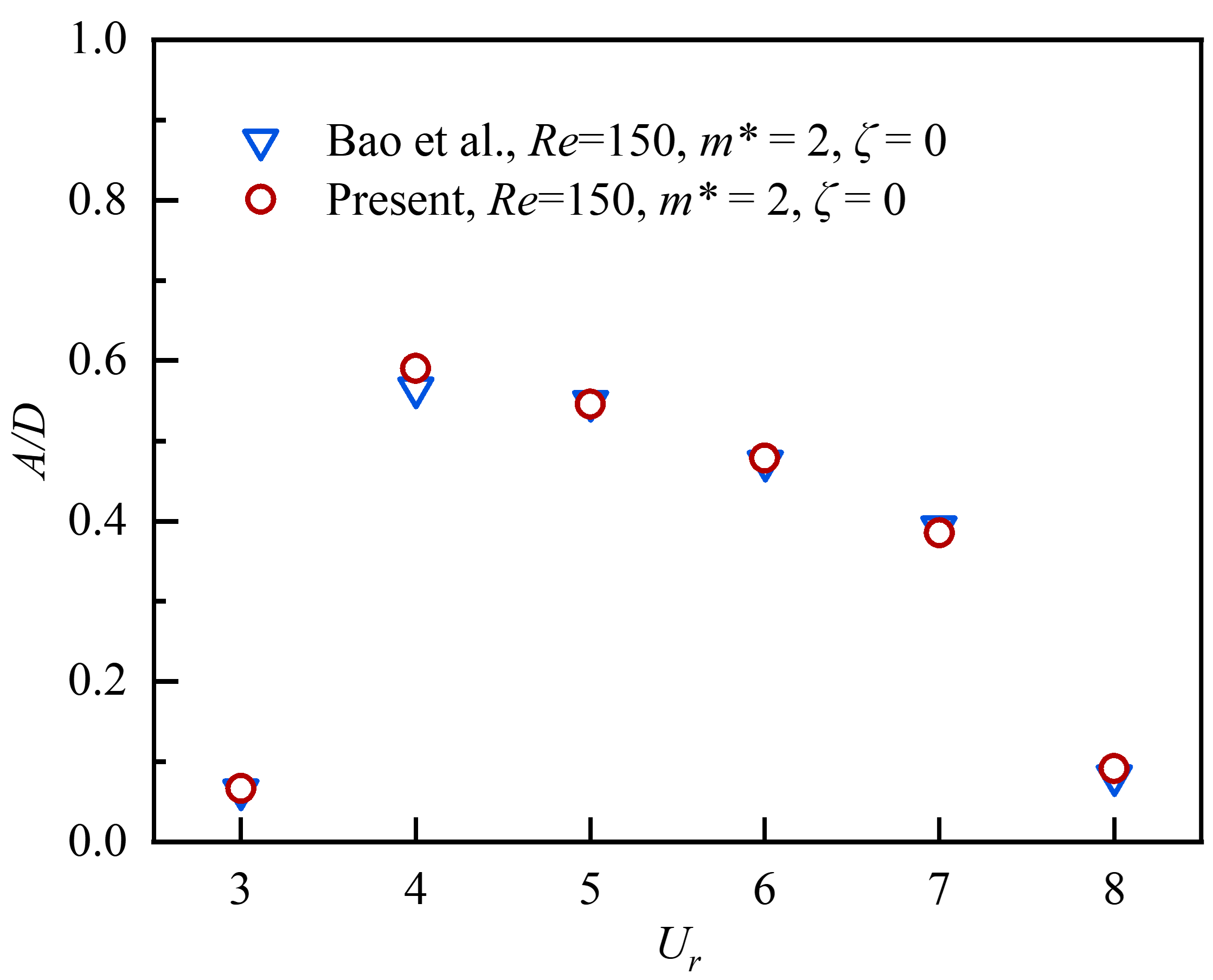}
    \caption{Validation of the fluid-structure interaction solver for the VIV of a single circular cylinder; here, $A/D$ is the non-dimensional amplitude of the vibration response.}
    \label{fig:fig4}
    \end{center}
\end{figure}

\hspace{0.5cm} With the chosen simulation setup, the results of the VIV of a single cylinder obtained using the present fluid-structure interaction solver are compared with the reference results of \citet{bao2012two} in the similar $Re$ regime to check its efficacy. Using the same parameters as \citet{bao2012two}, the comparison of dimensionless displacement of the cylinder is shown in Fig.~\ref{fig:fig4}. It can be seen that the present results are in good agreement with the reference results. Hence, the present solver can be used for the low $Re$ regime with confidence.

\section{Flow field characteristics}

In order to understand the effect of a downstream control cylinder, we systematically study the FIV characteristics of the primary cylinder and the underlying wake dynamics for a range of $Re$ and $L/D$ values. We first characterize the wake patterns through a detailed regime map and unravel the vortex interactions that underline the manifestation of different variety of wake patterns.

\subsection{Characterization of different Wake regimes}

\begin{figure}[htbp]
    \begin{center}
    \includegraphics[width=0.6\textwidth]{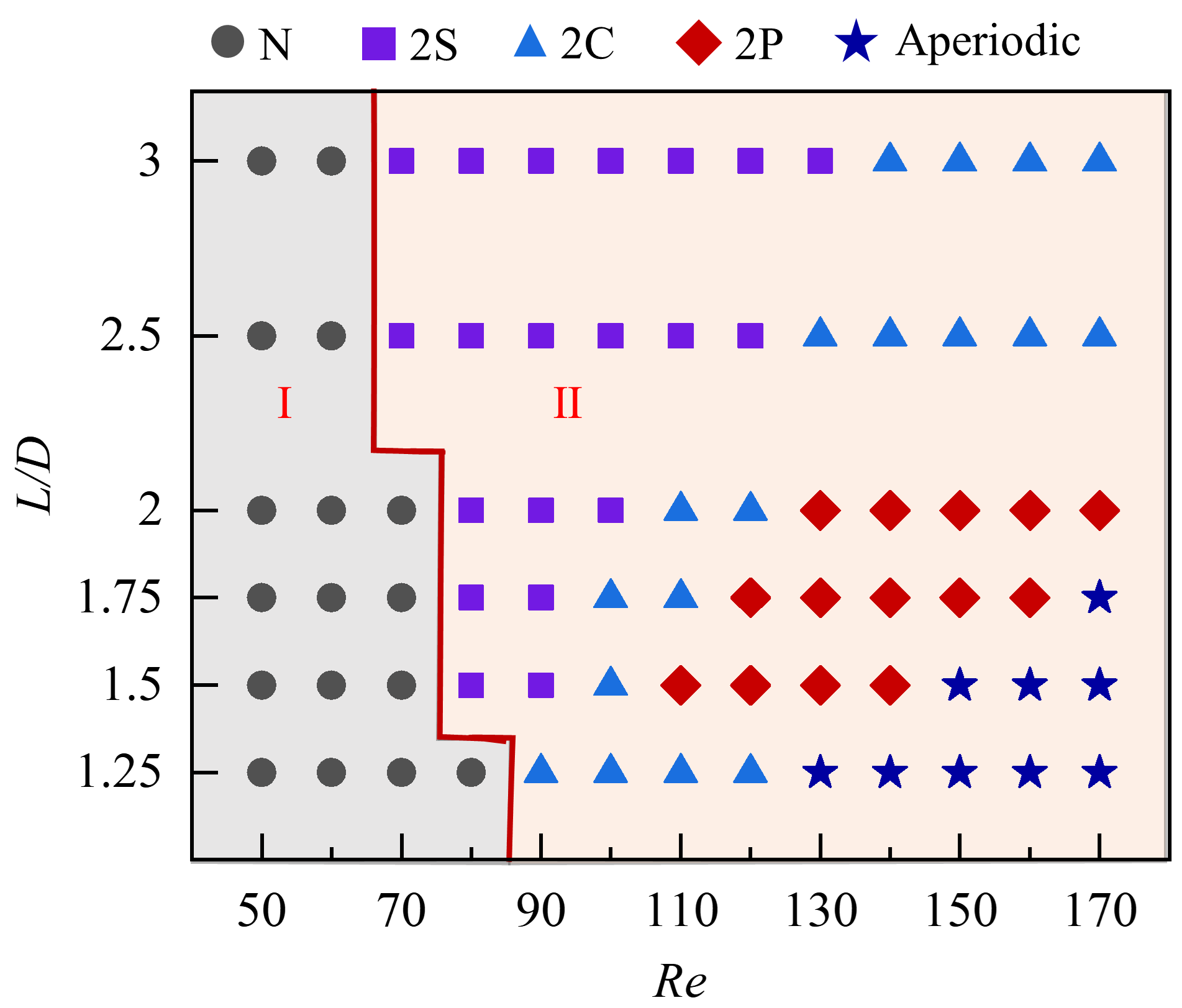}
    \caption{The wake regime map in $Re$–$L/D$ parametric space, where I represents the regime without vortex shedding, II represents the regime of alternate reattachment; the classification of wake based on the vortex shedding patterns: `N' (no vortex shedding), `2S', `2C', `2P', `Aperiodic'.}
    \label{fig:fig5}
    \end{center}
\end{figure}

Different wake regimes are characterized by varying $Re$ and $L/D$ values, and the wake regime map is presented in Fig.~\ref{fig:fig5}. The wake can be broadly distinguished into two different regimes according to the development of the shear layer \citep{ali2022heat}: the steady flow regime (denoted by I, where there is no vortex shedding behind the DC); the alternating attachment regime (denoted by II). As the $L/D$ value increases from 1.25 to 3, the range of $Re$, where steady flow occurs gradually decreases from $Re \le 80$ to $Re \le 60$, and no vortex shedding occurs in this region. On the other hand, the alternating attachment regime is predominant for $Re \ge 70$. The wake patterns are further classified according to the vortex shedding pattern of the two cylinders. As shown in Fig.~\ref{fig:fig5}, the alternating attachment regime can be divided into four sub-regions, namely (i) `2S' (two single isolated vortices are shed in an oscillation cycle of the cylinder), (ii) `2C' (two counter-rotating vortex pairs are shed in an oscillation cycle of the cylinder), (iii) `2P' (two co-rotating vortex pairs are shed in an oscillation cycle of the cylinder)\citep{stremler2011mathematical}, and (iv) aperiodic wake regime (vortex shedding takes place completely irregularly with no periodicity), respectively. It can be seen from the flow regime classification diagram that the characteristics of the flow field are sensitive to $Re$ and $L/D$. A detailed analysis of the flow field interactions is presented in the next subsection.

\subsection{Underlying vortex interactions}

\begin{figure}[htbp]
    \begin{center}
    \includegraphics[width=0.9\textwidth]{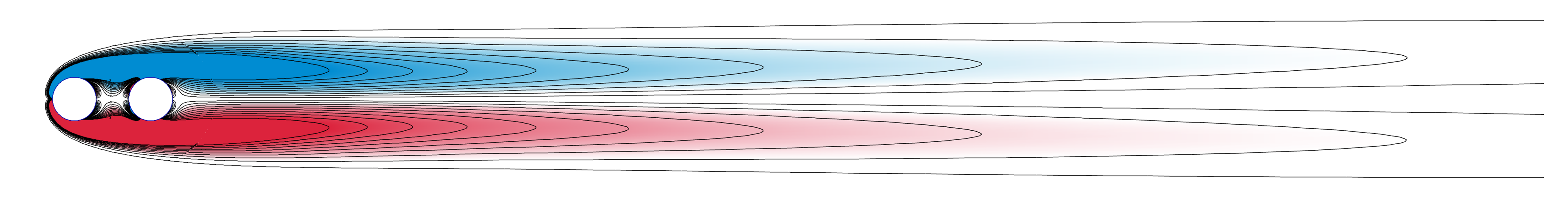} 
    \caption{The steady wake pattern ($-1 \le \omega_z^* \le 1$) at $L/D = 1.75, Re = 60$.}
    \label{fig:fig6}
    \end{center}
\end{figure}

\begin{figure}[htbp]
    \begin{center}
    \includegraphics[width=0.9\textwidth]{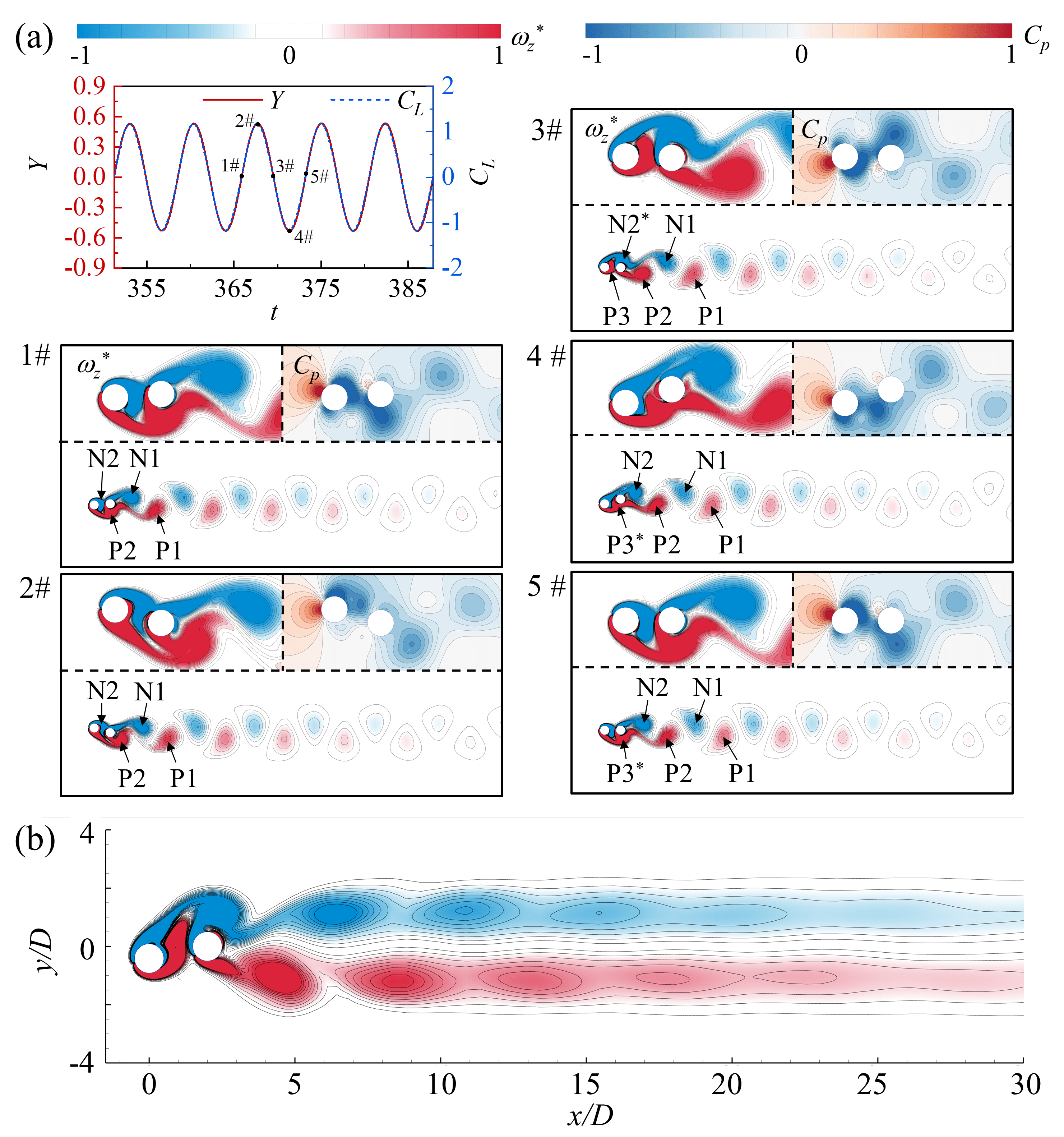} 
    \caption{The `2S' vortex shedding pattern ($-1 \le \omega_z^* \le 1$) at (a) $L/D = 1.75, Re = 80$; and (b) $L/D = 2, Re = 80$ and the corresponding time histories $Y$ and $C_L$ of the UC.}
    \label{fig:fig7}
    \end{center}
\end{figure}

The wake patterns for the steady flow regime and the `2S' vortex shedding pattern in the alternate reattachment regime are shown in Fig.~\ref{fig:fig6} and Fig.~\ref{fig:fig7}, respectively. The vorticity is non-dimensionalized, considering $U_{in}$ as the reference velocity scale and $D$ as the reference length scale, as $\omega_z^* = \omega_z D/ U_{in}$. In the steady flow regime (at $L/D$ = 1.75 and $Re$ = 60), the shear layer, separating from the UC, directly crosses over the DC, and no vortex shedding takes place behind the DC; see Fig.~\ref{fig:fig6}. The displacement ($Y$) and lift coefficient ($C_L$) of the UC are practically zero in this regime. Fig.~\ref{fig:fig7} shows the temporal evolution of the flow structures around the cylinders, the pressure coefficient ($C_P$) around the cylinders, and the corresponding time-histories of the $C_L$ and $Y$ (at $L/D$ = 1.75 and $Re$ = 80). Here, the clockwise vortices are marked in blue (negative), while the counter-clockwise vortices are marked in red (positive). The resulting wake pattern is characterized as a typical K\'arm\'an vortex street.

\hspace{0.5cm} At 1\# instance, the vortex P1 is seen to be separated from the shear layer of the DC, while the positive shear layer P2 generated by UC remains completely attached to the lower surface of DC. The negative vortex above the DC N1 is shed alternatively, and the negative shear layer N2 in the gap begins to roll up, adhering to the DC. As a result, the pressure on the upper surface of UC is lower than that on the lower surface, and the pressure difference of the DC is opposite to that of UC at this moment. At 2\# instance, as the UC reaches the maximum positive displacement, the negative shear layer in the gap starts adhering to the upper surface of DC. The shear layer P2 rolls up and squeezes the vortex N1. The pressure on the upper surface of UC increases significantly. At 3\# instance, the vortex N1 is separated, and the remaining shear layer merges with the negative shear layer N2, forming a new shear layer labelled N2$^*$. The negative pressure areas of UC and DC are seen to get reversed compared to the 1\# instance. The vortex shedding process at 3\# to 5\# instances is the same as that at 1\# to 3\# instances, but the vortex shedding direction is the opposite. In Fig.~\ref{fig:fig7}(a), the $C_L$ and the $Y$ time histories follow the same trend. Within the `2S' pattern sub-region, there is a distinctive wake pattern characterized by two-layered vortex streets, as shown in Fig.~\ref{fig:fig7}(b). This phenomenon is more pronounced at lower $Re$ and $L/D \ge 2$.

\begin{figure}[htbp]
    \begin{center}
    \includegraphics[width=0.9\textwidth]{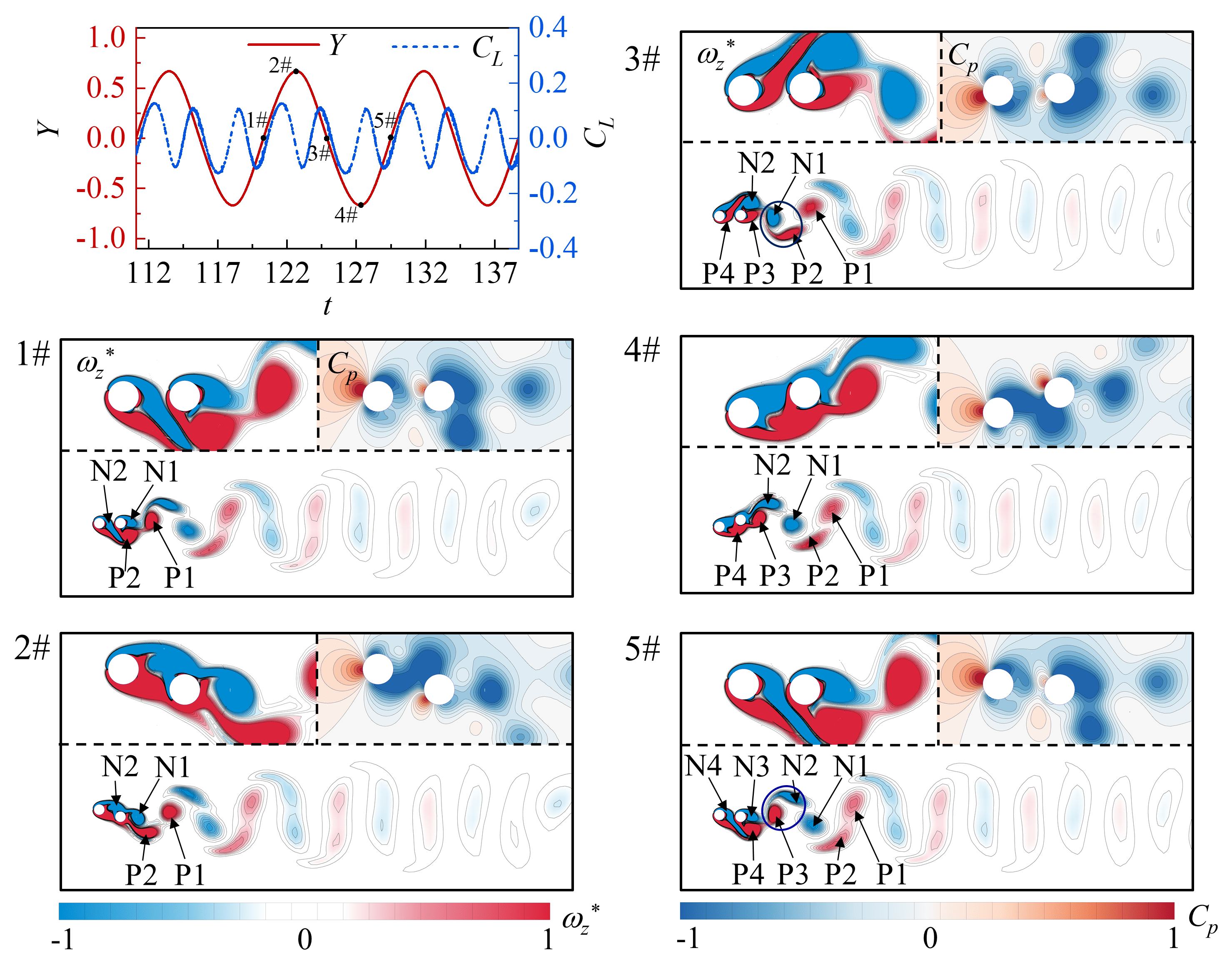} 
    \caption{The `2P' vortex shedding pattern ($L/D = 2, Re = 150$) in the alternate attachment regime and the corresponding time histories $Y$ and $C_L$ of the UC.}
    \label{fig:fig8}
    \end{center}
\end{figure}

\begin{figure}[htbp]
    \begin{center}
    \includegraphics[width=0.9\textwidth]{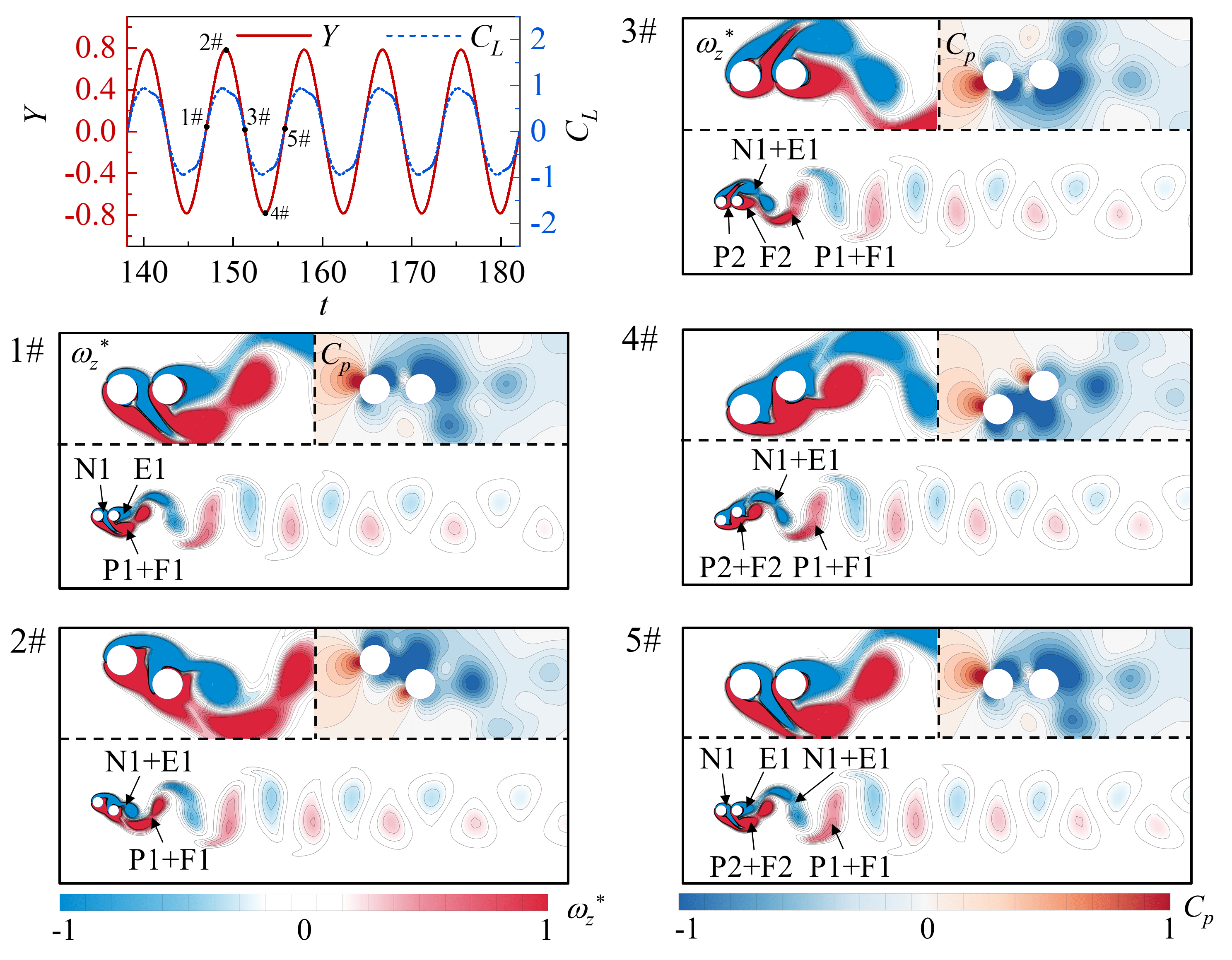} 
    \caption{The `2C' vortex shedding pattern ($L/D = 1.5, Re = 100$) in the alternate attachment regime and the corresponding time histories $Y$ and $C_L$ of the UC.}
    \label{fig:fig9}
    \end{center}
\end{figure}

\hspace{0.5cm} As shown in Fig.~\ref{fig:fig5}, the `2P' and `2C' vortex-shedding patterns occurs in a wide range of $Re$ and $L/D$. Figs.~\ref{fig:fig8} and \ref{fig:fig9} show the evolution process of `2P' and `2C' vortex-shedding patterns, the $C_P$ around the cylinders for different $Re$ and $L/D$ values, respectively. In Fig.~\ref{fig:fig8}, the cylinders are situated at the middle position at 1\# instance. The remaining shear layer from the negative shedding vortex of the previous cycle is marked as N1, while the negative shear layer within the gap is labeled as N2. Below the tandem cylinder, the positive shear layer from the UC has merged completely with that of the DC and is marked as P2. Both UC and DC have small pressure on the upper surface and high pressure on the lower surface. At 2\# instance, the UC reaches the positive maximum displacement. The downward-curling vortex N1 squeezes the shear layer P2, separating it from DC. On the upper surface of the DC, the shear layer N2 is fully attached to the DC, while the shear layers N1 and N2 remain separate. Conversely, the shear layer N2 promotes the shear layer N1 to separate from DC. At 3\# instance, the UC returns to the middle position. The shear layer P2 detaches from the DC due to the downward-curling shear layer N1. Due to the pushing and truncation effect of the shear layer N2 and P3, the shear layer N1 also sheds from DC. During the half cycle (from 3\# to 5\# instances) with negative motion displacement, the vortices N2 and P3 shed from the DC in a similar manner. In the downstream far-field of the DC, the vortices with the same direction between the two periods merge. The interference effect of DC results in a noticeable harmonic component of the frequency spectra of $C_L$, with the two peaks being almost equal.

\hspace{0.5cm} Figure~\ref{fig:fig9} illustrates the evolution process of the `2C' vortex shedding pattern. This vortex shedding pattern is observed between the `2S' and the `2P' or `aperiodic' wake regimes (when $L/D \le 2$) and on the right side of the `2S' pattern (when $L/D > 2$). At 1\# instance, the positive shear layers P1+F1 below DC are connected, and there are two shear layers, named E1 and N1, above DC and in the gap. At 2\# instance, the shear layer N1 attaches to the upper surface of DC and combines with E1 to form the new vortex named N1+E1. At time 3\#, vortex P1+F1 shedding from the DC. Compared with shear layer P3 at 3\# instance in Fig.~\ref{fig:fig8}, the truncation effect of the remaining shear layer F2 is weakened, and the combination of N1+E1 is not destroyed, but shedding together from the DC at 5\# instance in the negative half cycle. In the downstream far field of the DC, two vortices with the same direction merge completely into a vortex, the same as the typical K\'arm\'an vortex street. The evolution process of the `2S', `2C' and `2P' patterns reveals that with increasing $Re$, the vortex shedding pattern gradually transitions from the `2S' pattern to the `2P' pattern, while the downstream far-field will evolve to the typical K\'arm\'an vortex street.

\begin{figure}[htbp]
    \begin{center}
    \includegraphics[width=0.9\textwidth]{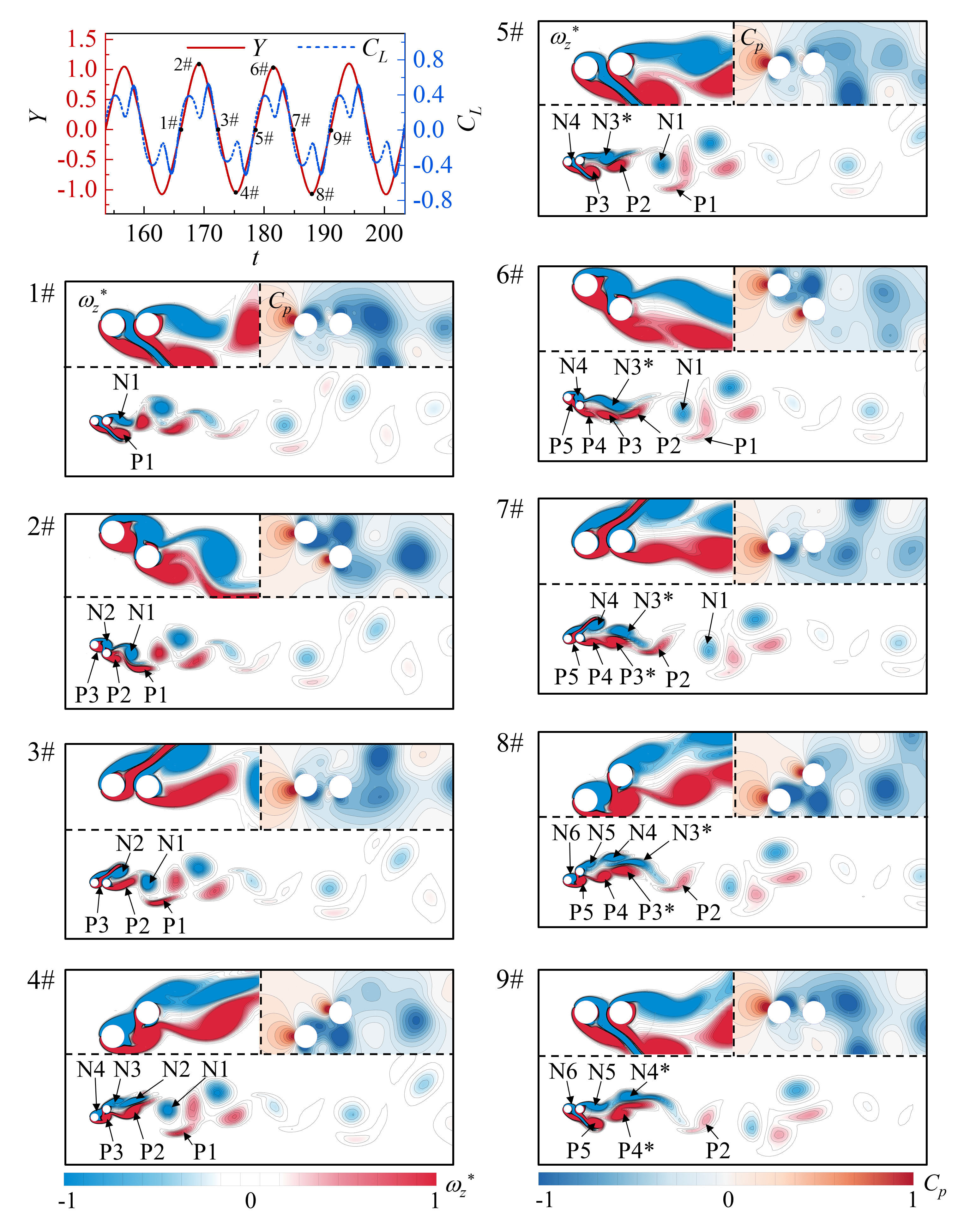} 
    \caption{The `aperiodic' vortex shedding pattern ($L/D = 1.5, Re = 160$) in the alternate attachment regime and the corresponding time histories $Y$ and $C_L$ of the UC.}
    \label{fig:fig10}
    \end{center}
\end{figure}

\begin{figure}[htbp]
    \begin{center}
    \includegraphics[width=0.9\textwidth]{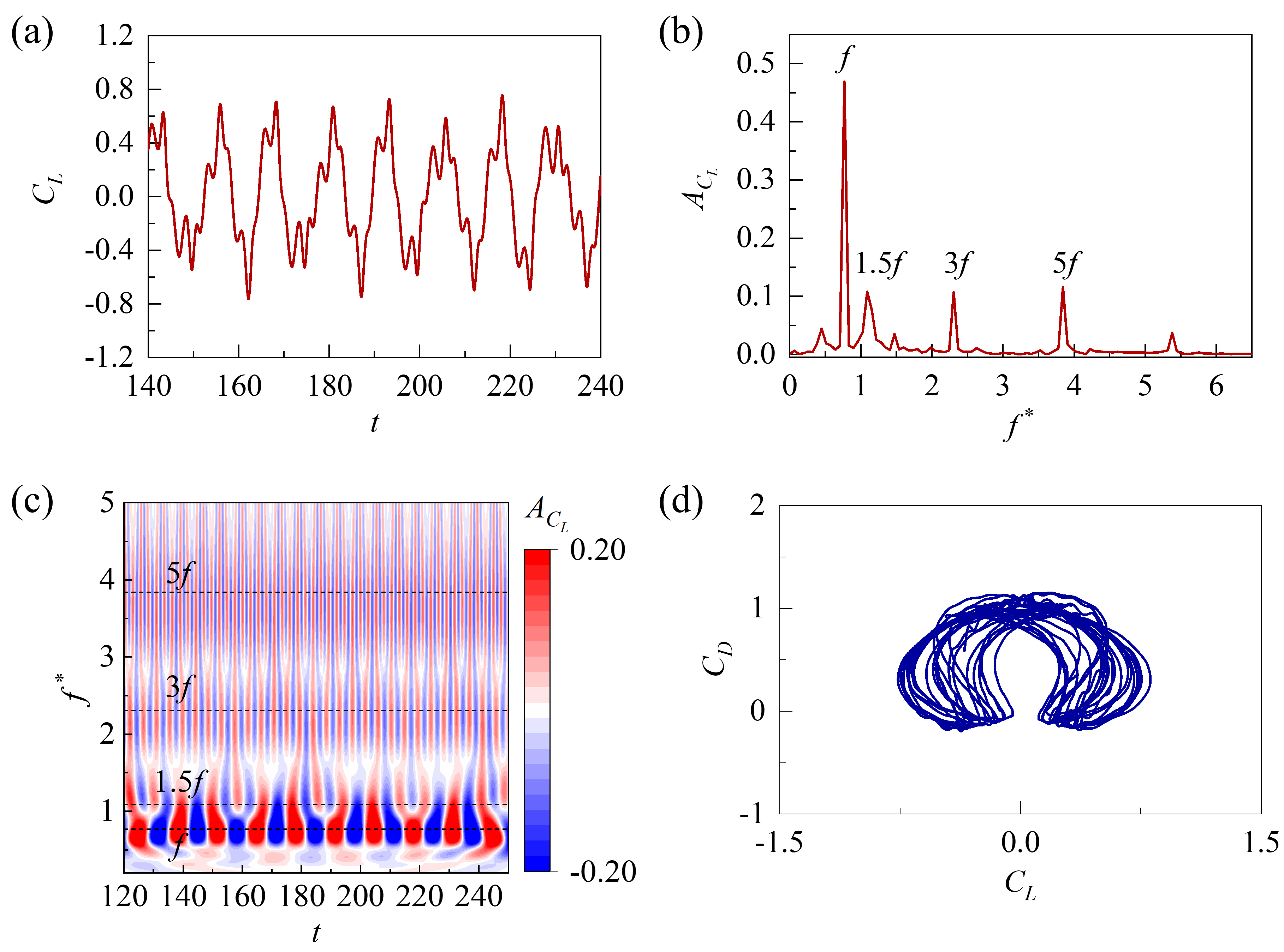} 
    \caption{Time series analysis of $C_L$ of DC for $L/D = 1.5, Re = 160$: (a) time history; (b) frequency spectra; (c) wavelet spectra; (d) $C_L-C_D$ phase portrait.}
    \label{fig:fig11}
    \end{center}
\end{figure}

\hspace{0.5cm} At small $L/D$ and high $Re$, the wake behind the tandem cylinders is characterized by an aperiodic vortex shedding pattern (`aperiodic' pattern), represented by a five-pointed star in the sub-region of Fig.~\ref{fig:fig5}. The corresponding time-histories of $C_L$ and $Y$ of the UC, the $C_P$ around the cylinders, and the evolution process of vortex shedding are shown in Fig.~\ref{fig:fig10}. It can be noted that the shedding pattern around the UC and the displacement of the UC are mostly periodic. However, the interference of the gap flow, and in turn, the wake downstream becomes aperiodic. At 1\# instance, vortices N1 and P1 are formed behind the DC. At 2\# instance, as the UC reaches its maximum positive displacement, the negative shear layer N2 is attached to the upper surface of DC. The positive shear layer below UC is truncated by DC, forming parts P2 and P3. Simultaneously, vortices N1 and P1 shed from the DC. At 3\# instance, shear layer P3 cuts vortex N2 into two parts, forming the shear layers N3 and N4 at the 4\# instance. It can be seen that at 4\# instance, the shear layer P3 has attached to the lower surface of DC, and the vortex P2 begins to separate from DC. At 5\# instance, the vortex P3 is divided into two parts by the shear layer N4. A new vortex labelled N3$^*$ emerges due to the merge of N2 and N3. At 6\# instance, the shear layer N4 attaches to the upper surface of DC, and N3$^*$ separates from the DC. Below the DC, the vortices P2, P3, and P4 form a long, narrow vortex strip. The vortex P3 merges with part of P4 to form a new vortex, P3$^*$, at the 7\# instance. The vortex N4 is divided into two parts by the shear layer P5 at 7\# instance and evolves into vortices N4, N5 and N6 at 8\# instance. Finally, the vortex N4 and N3$^*$ also merge to form N4$^*$ at 9\# instance, and the same is for vortex P4 and P3$^*$. The time-history, frequency spectra, wavelet spectra of $C_L$ for the DC, and the phase portraits of $C_L$ and $C_D$ of the DC at $L/D = 1.5$ and $Re = 160$ are presented in Figs.~\ref{fig:fig11}(a)-\ref{fig:fig11}(d), respectively. It can be seen that the time history of $C_L$ is irregular with broadband frequency spectra, indicative of the aperiodic nature of the flow field. It is evident that there is no distinct periodic attractor in the phase space. In comparison to the other three sub-regions in Fig.~\ref{fig:fig5}, this sub-region exhibits a complex process of wake vortices shedding and merging, lacking regularity in the vortex shedding. The instability of the shear layer separated from the UC can be attributed as the primary agency for triggering this aperiodicity over the DC. Hence, we classify this flow regime as an `aperiodic' wake pattern.


\subsection{Analysis of wake characteristics}
 The wake characteristics behind the tandem cylinder system are analyzed in this subsection. When the UC is in the equilibrium position just before the start of the upward movement, the flow field contour is considered to calculate the vortex width at a position of $x/D$ = 10 downstream. For the streamwise distance, the center-to-center distance between the two adjacent co-rotating vortex cores is calculated at $x/D$ = 10 downstream. All the cases are represented with the same contour levels. It is worth noting that, for the aperiodic cases, the average and standard deviation of the wake width and streamwise distance of the 10 consecutive oscillation cycles are calculated; the results are plotted with the corresponding error bars.

\begin{figure}[htbp]
    \begin{center}
    \includegraphics[width=1\textwidth]{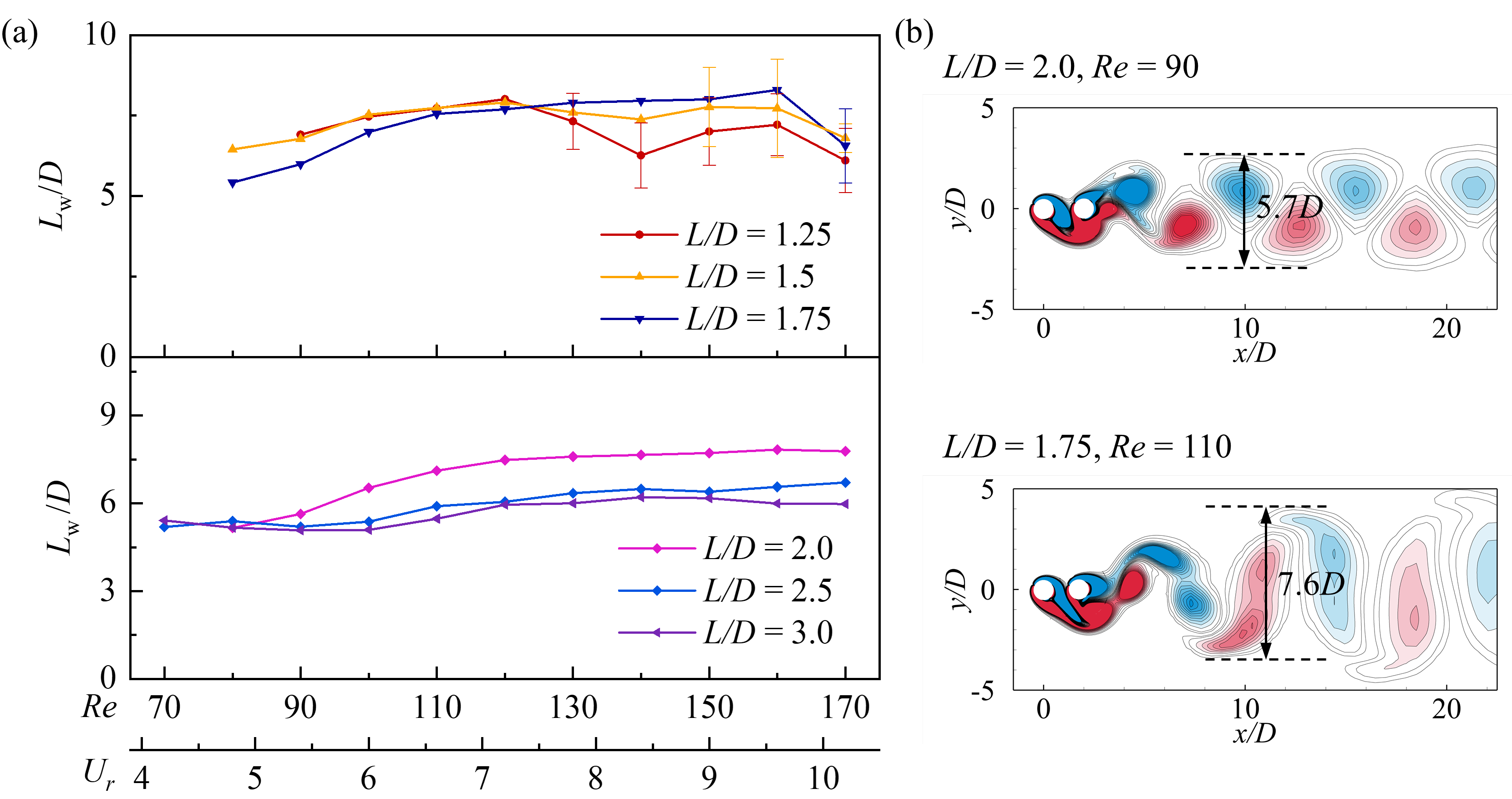} 
    \caption{(a) The variation of wake width at different $U_r$ and $Re$ values, (b) the details of two representative cases of vortex width calculations in the near-wake.}
    \label{fig:fig12}
    \end{center}
\end{figure}

Figure~\ref{fig:fig12}(a) illustrates the variation of the near-wake vortex width ($L_w/D$) for different $U_r$ and $Re$ values. The error bars in $L/D$ = 1.25, 1.5, and 1.75 represent the statistical results of the varying $L_w/D$ in the `aperiodic' pattern region. Two representative cases are presented in Fig.~\ref{fig:fig12}(b). At $L/D \le 2$, except for the `aperiodic' regime, the wake width gradually increases with the increase of reduced velocity, however, with a decreasing rate of increment. In the `aperiodic' regime, a variety of shedding patterns appear due to complex vortex shedding and merging; there is a different wake width in the wake flow. Comparing the vortex shedding pattern classification in Fig.\ref{fig:fig5}, the $L_w/D$ value increases rapidly in the `2S' and `2C' regimes, while only small changes are observed in the `2P' region. For $L/D$ values between 2.5 to 3 and $Re < 110$, the vortex shedding exhibits a `2S' pattern with a double-layered vortex street, and the varies of $L_w/D$ is small. Subsequently, as the vortex shedding transformed into `2S' pattern with the single-layer vortex street, the $L_w/D$ gradually increases.

\begin{figure}[htbp]
    \begin{center}
    \includegraphics[width=1.0\textwidth]{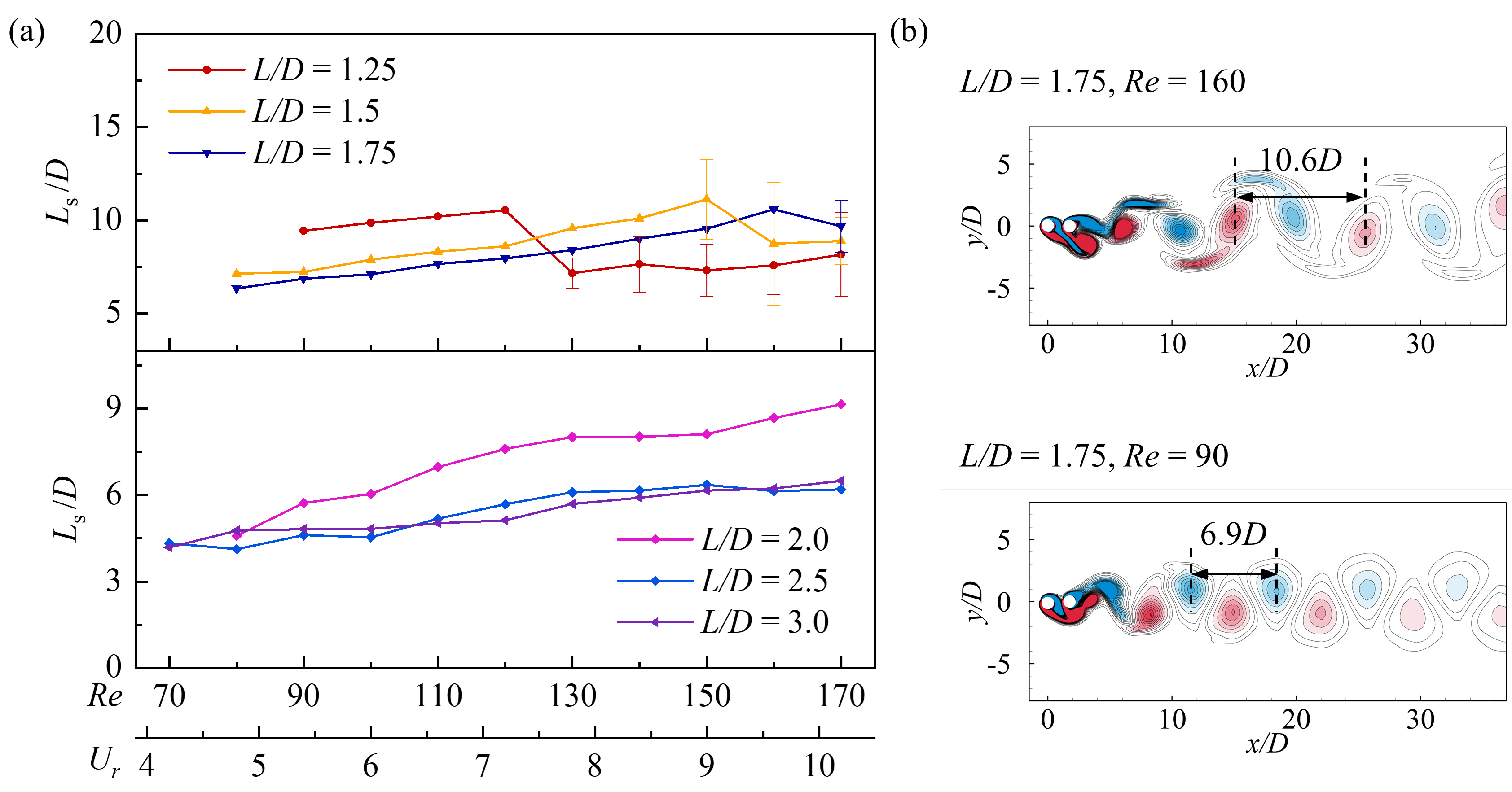} 
    \caption{(a) The variation of vortex spacing at different $U_r$ and $Re$ values, (b) the details of two representative cases of the streamwise vortex spacing calculation in the near-wake.}
    \label{fig:fig13}
    \end{center}
\end{figure}

\hspace{0.5cm} Figure~\ref{fig:fig13}(a) presents streamwise distance ($L_s/D$) for the two adjacent co-rotating near-wake vortices as a function of $U_r$ and $Re$. Two representative cases are presented in Fig.~\ref{fig:fig13}(b). Different $L_s/D$ values are observed in the consecutive vortex shedding periods in the `aperiodic' wake regime, which can be attributed to the complex vortex shedding behind the DC. When $L/D$ ranges from 2.5 to 3, $L_s/D$ increases with increasing $U_r$, indicating a decrease in the vortex shedding frequency. Based on the vortex shedding pattern classification in Fig.~\ref{fig:fig5}, the $L_s/D$ gradually increases as the vortex shedding pattern evolves from `2S' to `2C' and `2P'. When located in the two-layered vortex street, the change of $L_s/D$ is small. 

\section{Flow-induced vibration characteristics}

In the previous section, the different wake regimes observed with different $Re$ are analyzed. The results have shown that a strong interference effect due to the presence of the DC affects the vortex shedding patterns of the UC, in turn significantly altering the associated aerodynamic forces. Next, in this section, the flow-induced vibration characteristics of the UC are presented.

\subsection{Amplitude and frequency characteristics}

\begin{figure}[htbp]
    \begin{center}
    \includegraphics[width=1.0\textwidth]{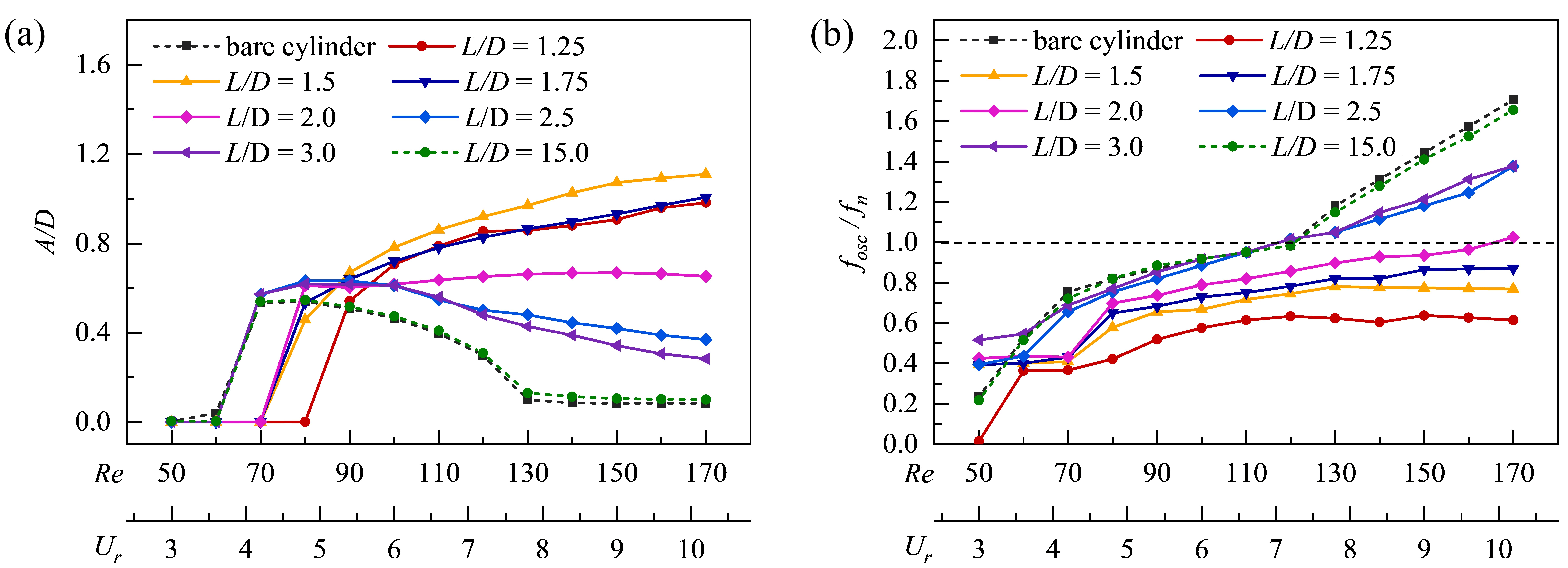} 
    \caption{The variations in (a) the dimensionless amplitude ($A/D$) of the FIV response of the UC and (b) the corresponding dimensionless frequency ratio ($f_{osc}/f_n$) with $Re$.}
    \label{fig:fig14}
    \end{center}
\end{figure}

\begin{figure}[htbp]
    \begin{center}
    \includegraphics[width=0.6\textwidth]{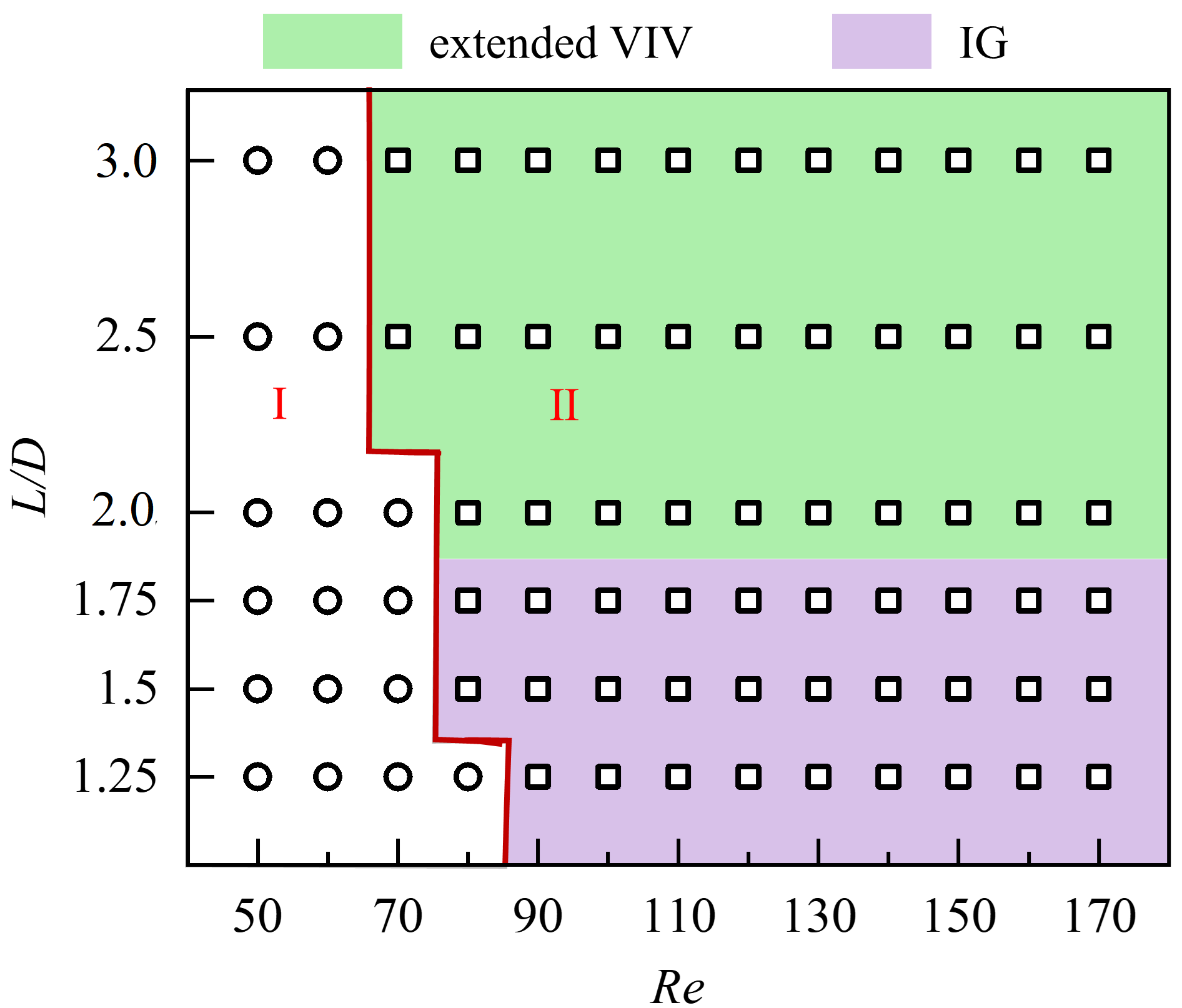} 
    \caption{Dynamical classification of the vibration response, where circle refers to region I, square refers to region II, and pentagon refers to region III.}
    \label{fig:fig15}
    \end{center}
\end{figure}

Figure~\ref{fig:fig14} shows the variation of the dimensionless amplitude ($A/D$) and the frequency ratio ($f_{osc}/f_n$) of the flow-induced vibration response of the UC as a function of $U_r$, in comparison to the results of a single cylinder and large spacing ratio $L/D = 15$. Here, the amplitude $A$ is defined as $0.5(y_{max}-y_{min})$, where $y_{max}$ and $y_{min}$ are the maximum and minimum response amplitude, respectively. Figure~\ref{fig:fig15} presents the dynamical classification of the vibration response, in which I represent the steady flow regime, where the displacement of the UC is close to 0; II is the alternating attachment regime. Two vibration types of the UC are distinguished by colour blocks, extended VIV and interference galloping (IG). As can be seen in Fig.~\ref{fig:fig14}(a), the dimensionless amplitude of the UC increases gradually with increasing $U_r$ and has a similar trend when $L/D$ varies between 1.25 and 1.75. In addition, the amplitude is the largest when the $L/D = 1.5$. When $L/D \ge 2.5$ and $U_r \ge 5.4$, the dimensionless amplitude shows a decreasing trend with increasing $L/D$; the amplitude does not change significantly for $L/D = 2$. It can be seen from Fig.~\ref{fig:fig14}(b) that the dimensionless frequency of the UC gradually increases with increasing $U_r$ after the UC starts vibrating. Besides, the dimensionless frequency also increases with increasing $L/D$.


\hspace{0.5cm} When $L/D$ is 1.25, 1.5, and 1.75, the dimensionless frequencies are locked-in at 0.62, 0.77, and 0.87, respectively, and the vibration response of the UC can be characterized as the interference galloping \citep{dielen1995mechanism}. The interference effect of the DC on the vibration of the UC gradually weakens with the increasing $L/D$. When $L/D$ is greater than a critical threshold, the response of the UC transitions to pure VIV \citep{chunning2014numerical}. As can be seen from Figs.~\ref{fig:fig14}(a) and~\ref{fig:fig14}(b), when $L/D = 15$, the dimensionless amplitude and the frequency ratio of the UC response are completely consistent with that of a single circular cylinder. In Fig.~\ref{fig:fig14}(a), when $L/D$ is between 2.5 and 3, the dimensionless amplitude reaches its maximum at $U_r$ = 4.8, and then gradually decreases with the increasing $U_r$. It is worth noting that, when $U_r \ge 7.8$, the dimensionless displacement of the UC is larger than that of a single cylinder. The dimensionless frequency of the UC increases from 0.68 to 1.37 when $L/D$ is between 2.5 and 3. When $L/D$ = 2, the maximum frequency ratio is 1.02 at $U_r$ = 10.2, which is smaller than that observed for the $L/D$ between 2.5 to 3. This indicates that due to the interference effect of the DC, there is a wider lock-in region when $L/D$= 2. According to \citet{chen2022flow}, when $L/D$ falls between 2 and 3, it can be classified as extended VIV. The range of $L/D$ also aligns with \citet{bokaian1984proximity}.

\hspace{0.5cm} According to the classification results of the wake regime presented in Fig.~\ref{fig:fig15}, there is no vortex shedding behind the cylinder in the steady flow regime, and the displacement of the UC is close to zero in this regime. As $U_r$ increases, the wake changes from the steady flow to the alternate attachment regime, and the UC begins to vibrate with a large amplitude. Comparing with the classification in Fig.~\ref{fig:fig5}, the wake shifts from steady flow to the `2S', and then evolves into `2C', `2P', and `aperiodic' pattern with increasing $U_r$ at $L/D$ = 1.5, is different from the evolution process from `2C' to `aperiodic' pattern at $L/D$ = 1.25, resulting a small starting velocity and larger vibration amplitude of the former. At $L/D$ = 1.75, the larger transition range of the `2P' pattern and smaller distribution of the `aperiodic' pattern results in smaller vibration amplitudes compared with $L/D$ = 1.5.

\subsection{Phase difference characteristics}

\begin{figure}[htbp]
    \begin{center}
    \includegraphics[width=0.6\textwidth]{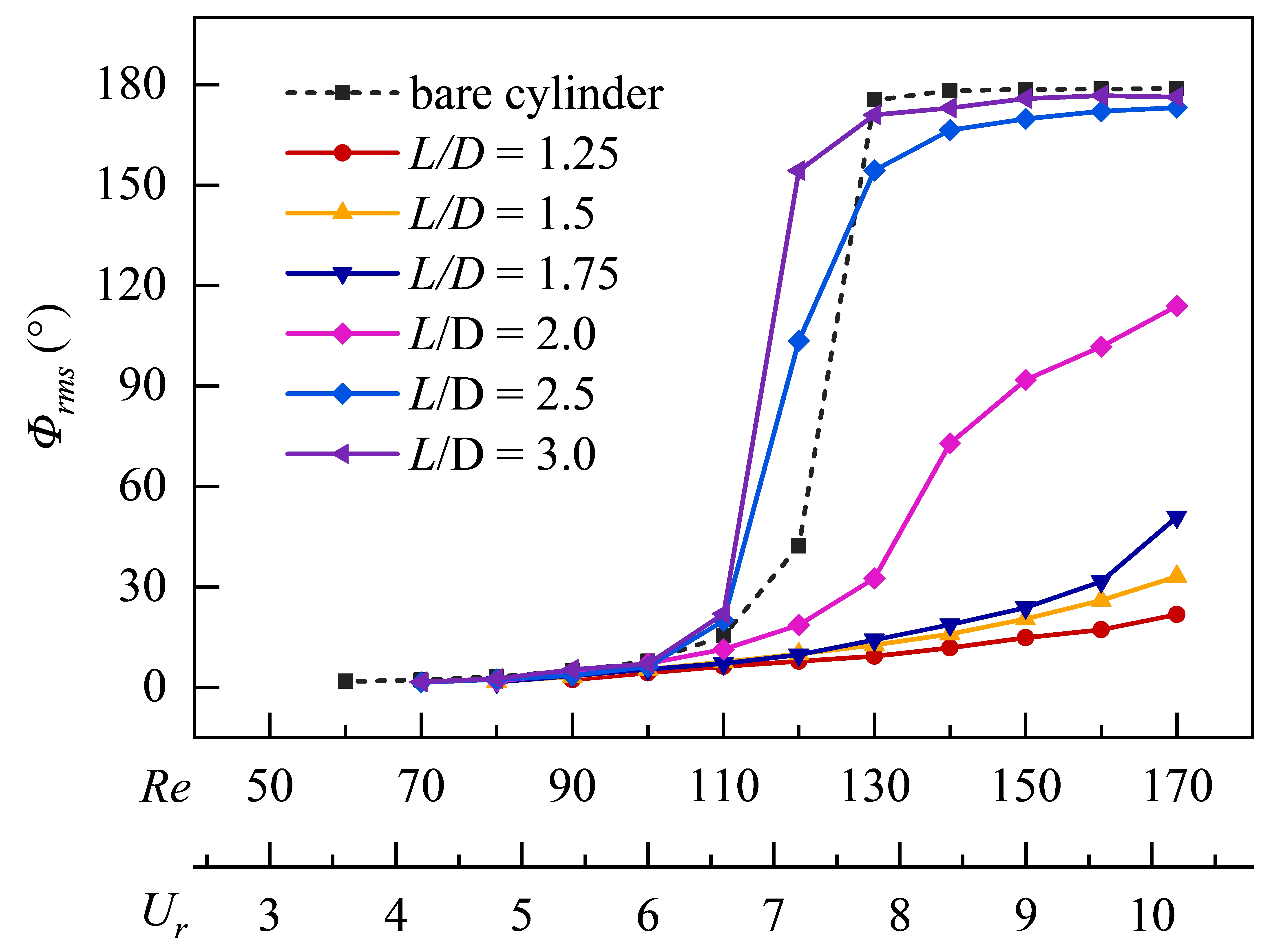} 
    \caption{Variation of the phase difference between lift force and transverse displacement with $Re$.}
    \label{fig:fig16}
    \end{center}
\end{figure}

Using the Hilbert transform, the instantaneous phase of the lift coefficient and the dimensionless amplitude is calculated, and then the instantaneous phase difference between the two is obtained. Finally, the root mean square value of the instantaneous phase difference ($\phi_{rms}$) is estimated to evaluate the phase characteristics of the lift coefficient and the dimensionless amplitude; see Fig.~\ref{fig:fig16}.
When $L/D \le 1.75$, the $\phi_{rms}$ gradually increases with the increasing $U_r$. The $\phi_{rms}$ increases to $51^\circ$ when $U_r = 10.2$ and $L/D = 1.75$, and increases to $110^\circ$ when $L/D$ increases to 2. Consequently, the increase in the dimensionless displacement decreases with increasing $U_r$. At $L/D$ = 2, the $\phi_{rms}$ increases significantly, and the enhancing effect of lift on the cylinder vibration weakens, resulting in a small change in the amplitude of the UC. The $\phi_{rms}$ of the UC are similar to that of the single circular cylinder for $L/D \ge 2.5$. When $U_r > 7.8$ and $L/D \ge 2.5$, the $\phi_{rms}$ reaches $180^\circ$ (the lift coefficient and the dimensionless displacement are in anti-phase condition), resulting in continuous decreases of the response amplitude of the UC with increasing $U_r$.

\section{Force characteristics of the upstream cylinder}

\begin{figure}[htbp]
    \begin{center}
    \includegraphics[width=1.0\textwidth]{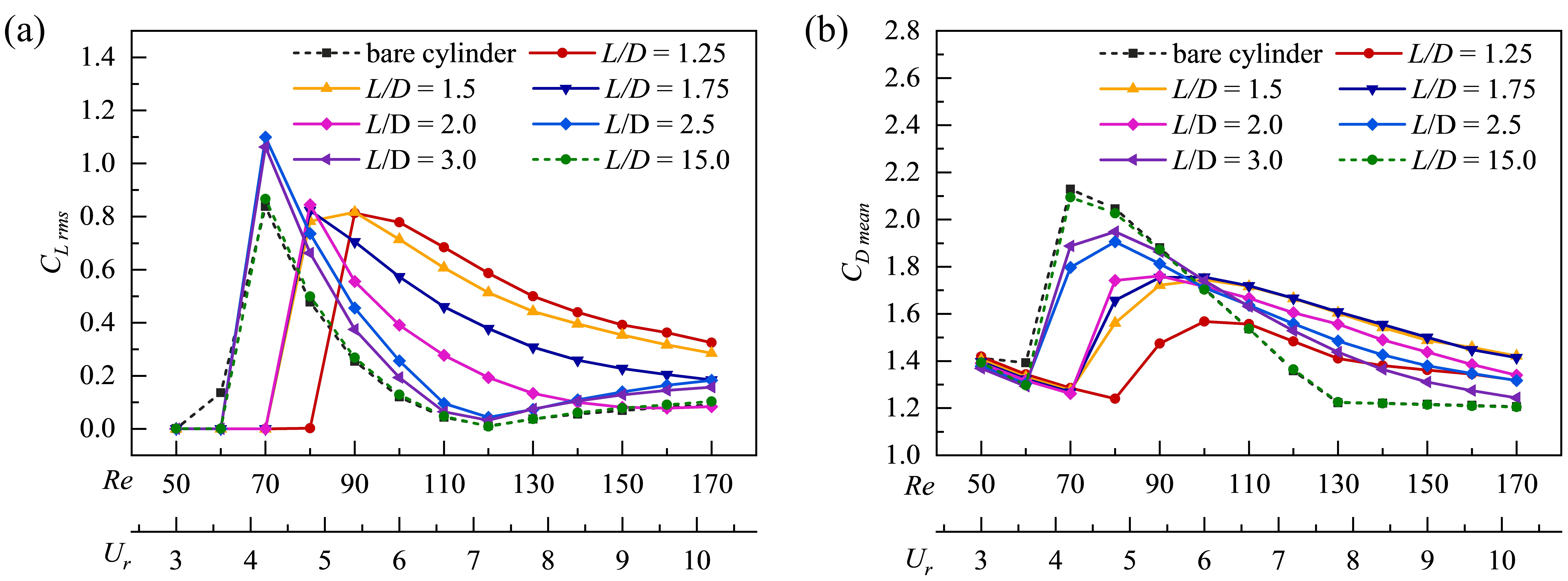}
    \caption{The variation of (a) ${C_L}_{rms}$ and (b) ${C_D}_{mean}$ with $Re$ and $U_r$ for various $L/D$ values.}
    \label{fig:fig18}
    \end{center}
\end{figure}

Figure~\ref{fig:fig18} shows the variation of the root mean square value of the lift coefficient (${C_L}_{rms}$) and the mean value of the drag coefficient (${C_D}_{mean}$) of the UC with increasing $Re$ and $U_r$. It can be seen from Fig.~\ref{fig:fig18}(a) that the ${C_L}_{rms}$ is almost zero at the steady flow regime due to the absence of vortex shedding. As the $U_r$ increases, the ${C_L}_{rms}$ undergoes a sudden jump with the change in flow regime. When $L/D = 1.25$ and $Re$ between 80 to 90 or $L/D = 1.5$ and $Re$ between 90 to 100, the change of the ${C_L}_{rms}$ is very small in the alternate attachment regime, and then gradually decreases with the increasing $U_r$. On the contrary, the ${C_L}_{rms}$ of the UC jumps to the maximum along with the change of flow regime and then decreases with increasing $U_r$ at the condition of $L/D \ge 1.75$. It is worth noting that the ${C_L}_{rms}$ slightly increases when $U_r \ge 7.2$ and $L/D \ge 2.5$ compared to the single cylinder case. This coincides with a slow decreasing trend of the response displacement of the UC. It can be seen from Fig.~\ref{fig:fig18}(b) that the ${C_D}_{mean}$ value for different $L/D$ follows the same trend. There is a small decrease in the steady flow regime. Then, it increases significantly after the onset of the UC vibration and subsequently decreases with increasing $Re$. The ${C_D}_{mean}$ of the UC is larger than that of a single cylinder when $U_r > 6.6$. It is worth noting that the critical $Re$ for steady to unsteady flow transition decreases as $L/D$ increases. At $L/D = 15$, the interference effect of the DC on the UC disappears, and the UC and the single circular cylinder are seen to have the same ${C_L}_{rms}$ and ${C_D}_{mean}$.

\begin{figure}[htbp]
    \begin{center}
    \includegraphics[width=1.0\textwidth]{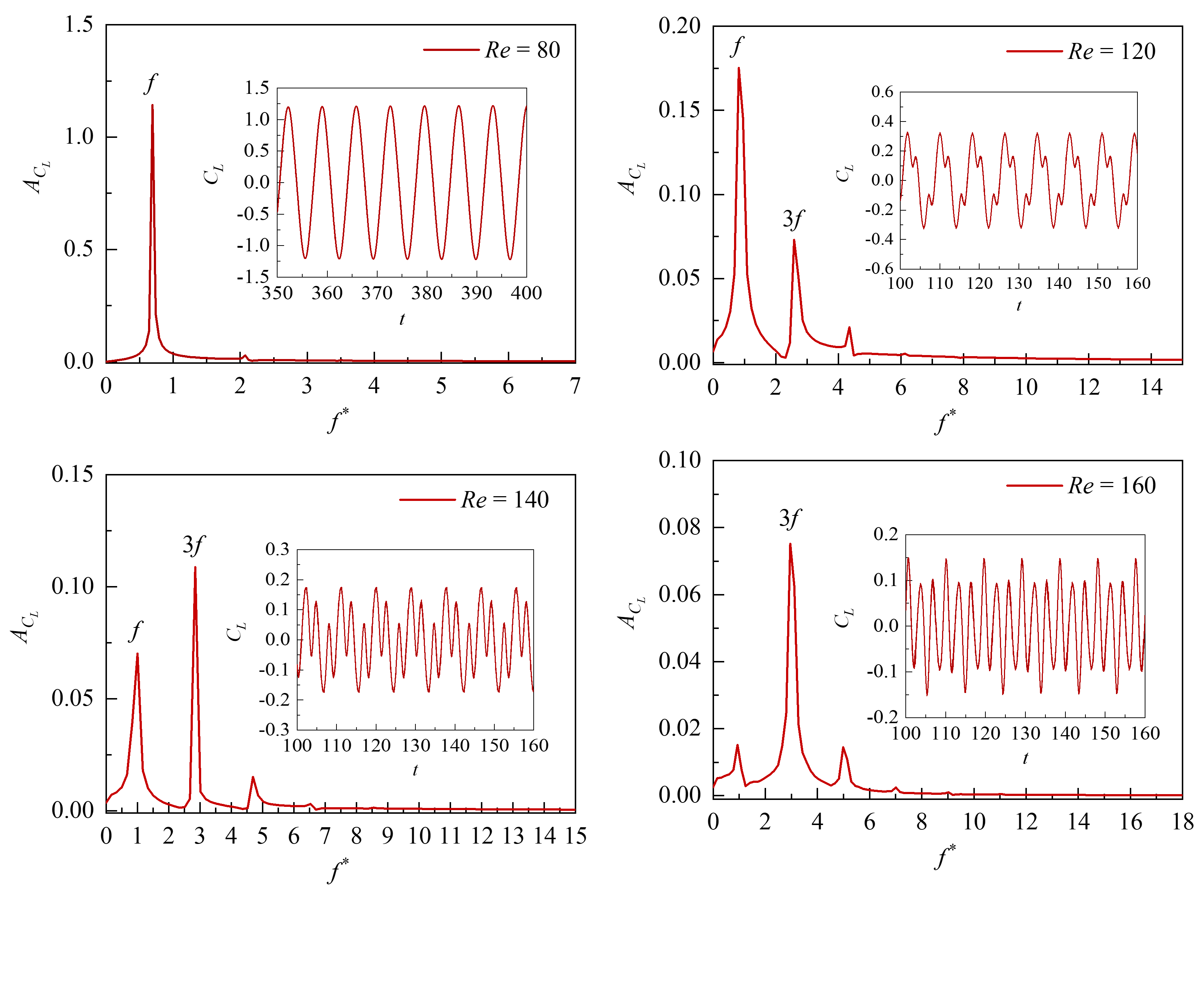}
    \caption{The time history and frequency spectra of the $C_L$ for $L/D = 2$ and different $Re$.}
    \label{fig:fig19}
    \end{center}
\end{figure}

\begin{figure}[htbp]
    \begin{center}
    \includegraphics[width=1.0\textwidth]{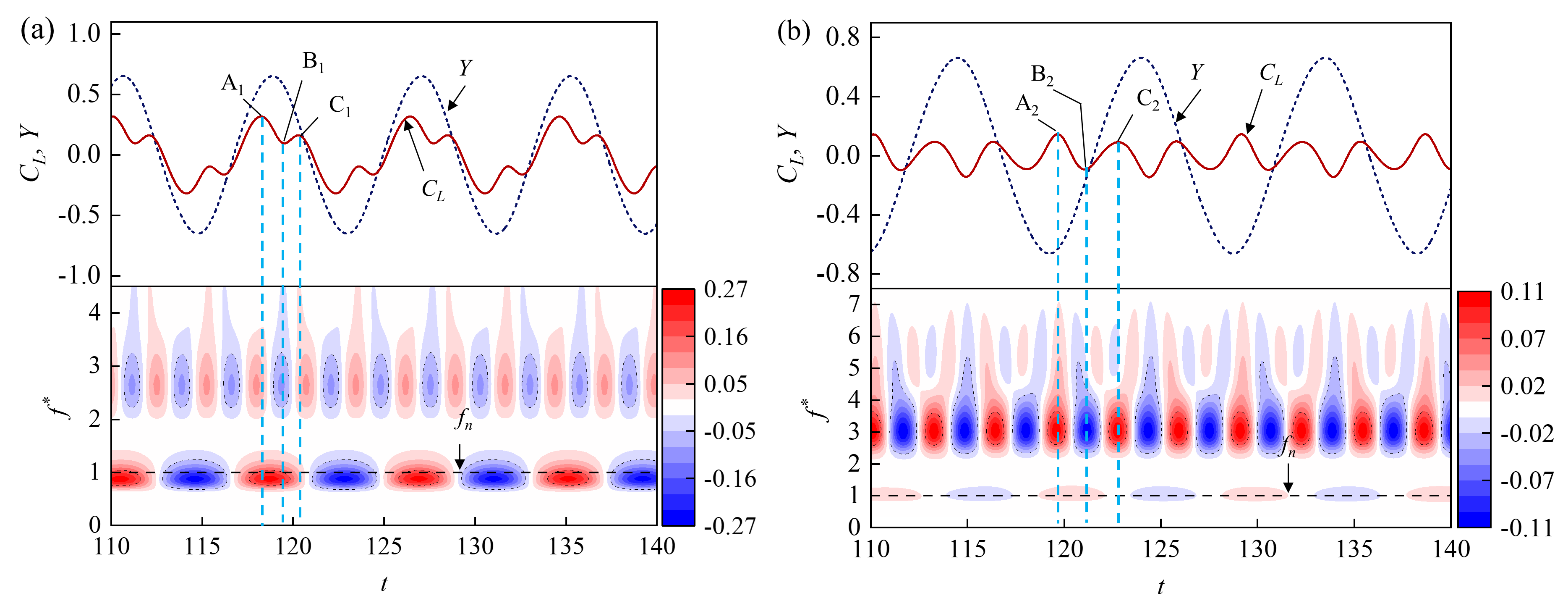}
    \caption{The continuous wavelet transform (CWT) results of the $C_L$ time histories: (a) $L/D = 2$, $Re = 120$; (b) $L/D = 2$, $Re = 160$.}
    \label{fig:fig20}
    \end{center}
\end{figure}

\hspace{0.5cm} The time-histories and frequency spectra of $C_L$ for $L/D = 2$ and different $Re$ are shown in Fig.~\ref{fig:fig19}. It can be seen that the frequency spectra primarily comprise a dominant frequency peak (vortex shedding frequency) and its triple harmonic. Due to the interference effect of the DC on the vortex shedding, the triple frequency harmonic component on the time history of $C_L$ becomes more significant as $Re$ increases. Notably, at $Re$ = 160, the magnitude of the triple harmonic component exceeds the magnitude of the fundamental frequency and becomes the dominant shedding frequency. The earlier studies \citep{wu2012review} showed that the triple harmonic lift is related to three vortices shedding behind the cylinder, resulting in a `2T' vortex shedding pattern. For the present tandem cylinder system, although the `2T' vortex shedding pattern does not appear, the triple harmonic frequency of $C_L$ of UC can be attributed to the shear layer interaction in the intermediate gap flow. The wavelet analysis of the $C_L$ time-history is carried out for $Re$ = 120 and $Re$ = 160 at $L/D$ = 2; see Fig.~\ref{fig:fig20}. The contour represents the variation of the Continuous Wavelet Transform coefficient with non-dimensional time. The black dotted line is the natural frequency ($f_n$) of the elastically supported cylindrical structure. From the contour in Fig.~\ref{fig:fig20}(a), it can be seen that when the $Re$ = 120, the peaks A1 and C1 at the positive half period of the lift coefficient are located between the positive and negative peaks of the triple harmonic component. 
When $Re$ = 120, the two peaks are mainly affected by the first-order fundamental component. 
While $Re$ = 160, the variation of the lift coefficient is similar to that of $Re$ = 120; see Fig.~\ref{fig:fig20}(b). However, the peaks (A2 and C2) of the lift coefficient coincide with the peak of the triple harmonic component, that is, the lift coefficient is mainly dominated by the triple frequency harmonic lift.

\begin{figure}[htbp]
    \begin{center}
    \includegraphics[width=0.9\textwidth]{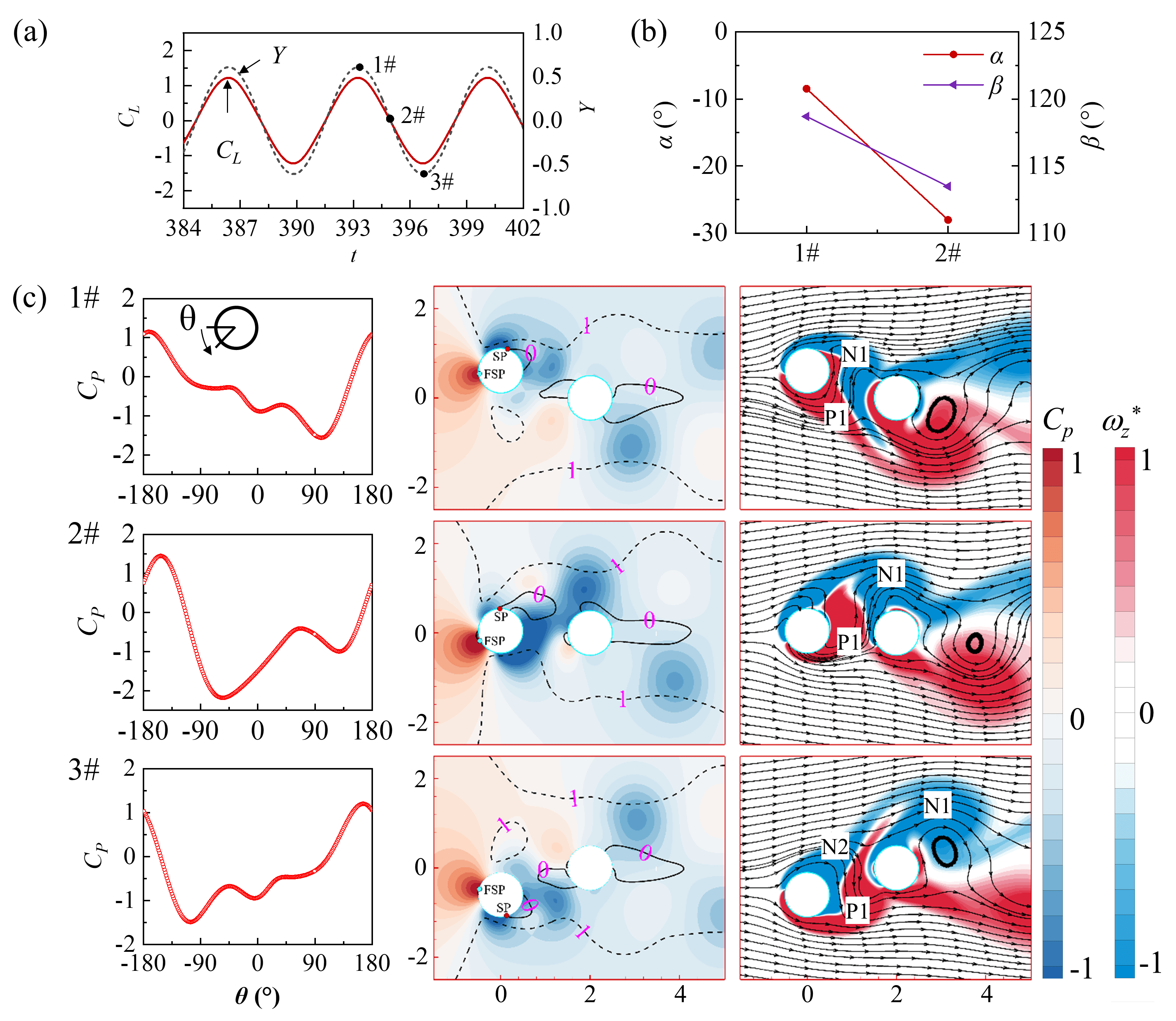}
    \caption{(a) lift coefficient and dimensionless amplitude, (b) the angle between the front stagnation point and the separation point of the boundary layer at 1\# moment, and (c) the surface pressure coefficient and the contour of the surface pressure coefficient and the local vorticity at different times when $Re = 80$.}
    \label{fig:fig21}
    \end{center}
\end{figure}

\begin{figure}[htbp]
    \begin{center}
    \includegraphics[width=0.80\textwidth]{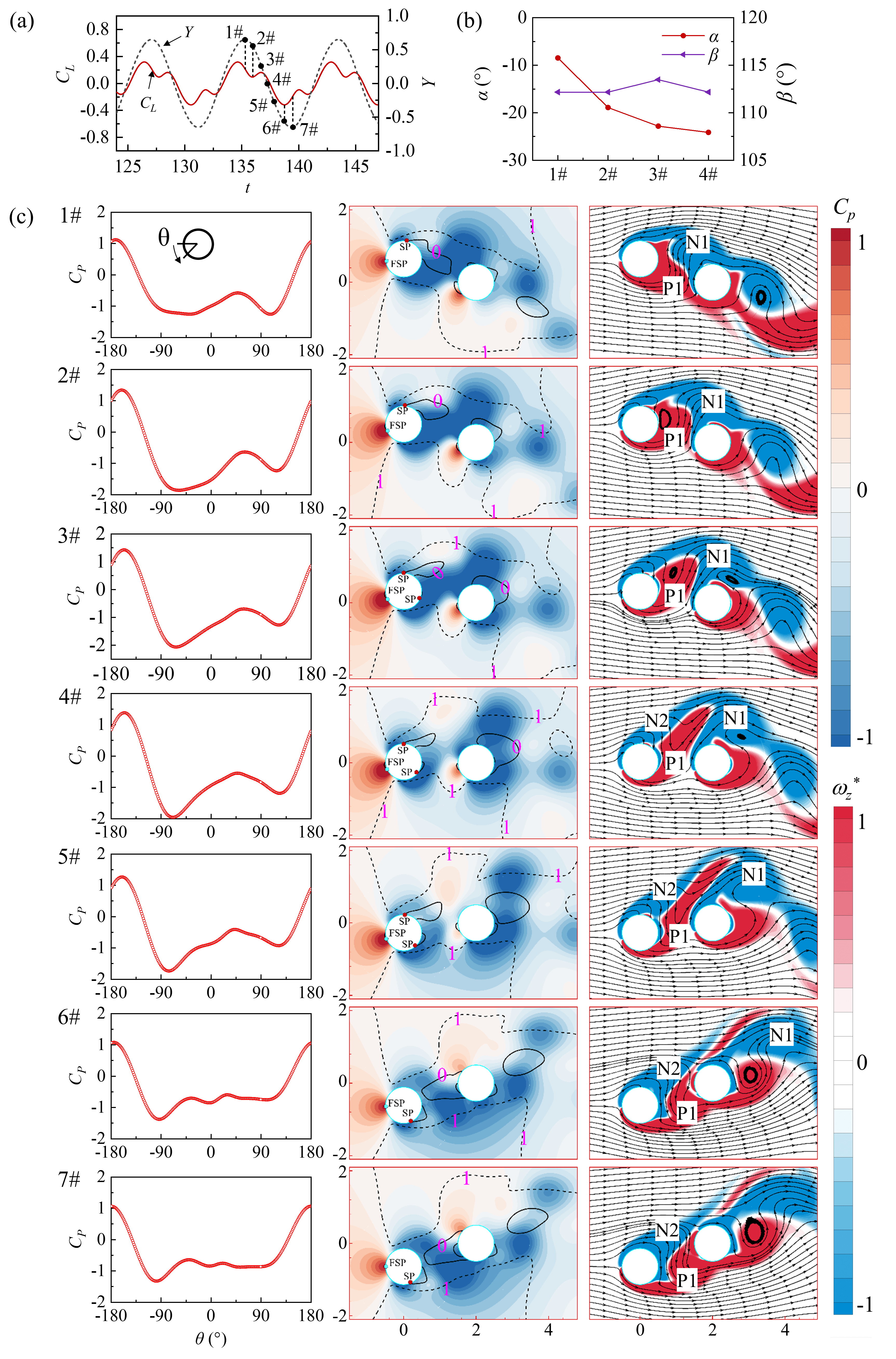}
    \caption{(a) lift coefficient and dimensionless amplitude, (b) variations of the $\alpha$ and $\beta$ in a half cycle of the displacement from 1\# to 4\#, (c) the surface pressure coefficient and the contour of the surface pressure coefficient and the local vorticity at different times when $Re = 120$.}
    \label{fig:fig22}
    \end{center}
\end{figure}

\begin{figure}[htbp]
    \begin{center}
    \includegraphics[width=0.80\textwidth]{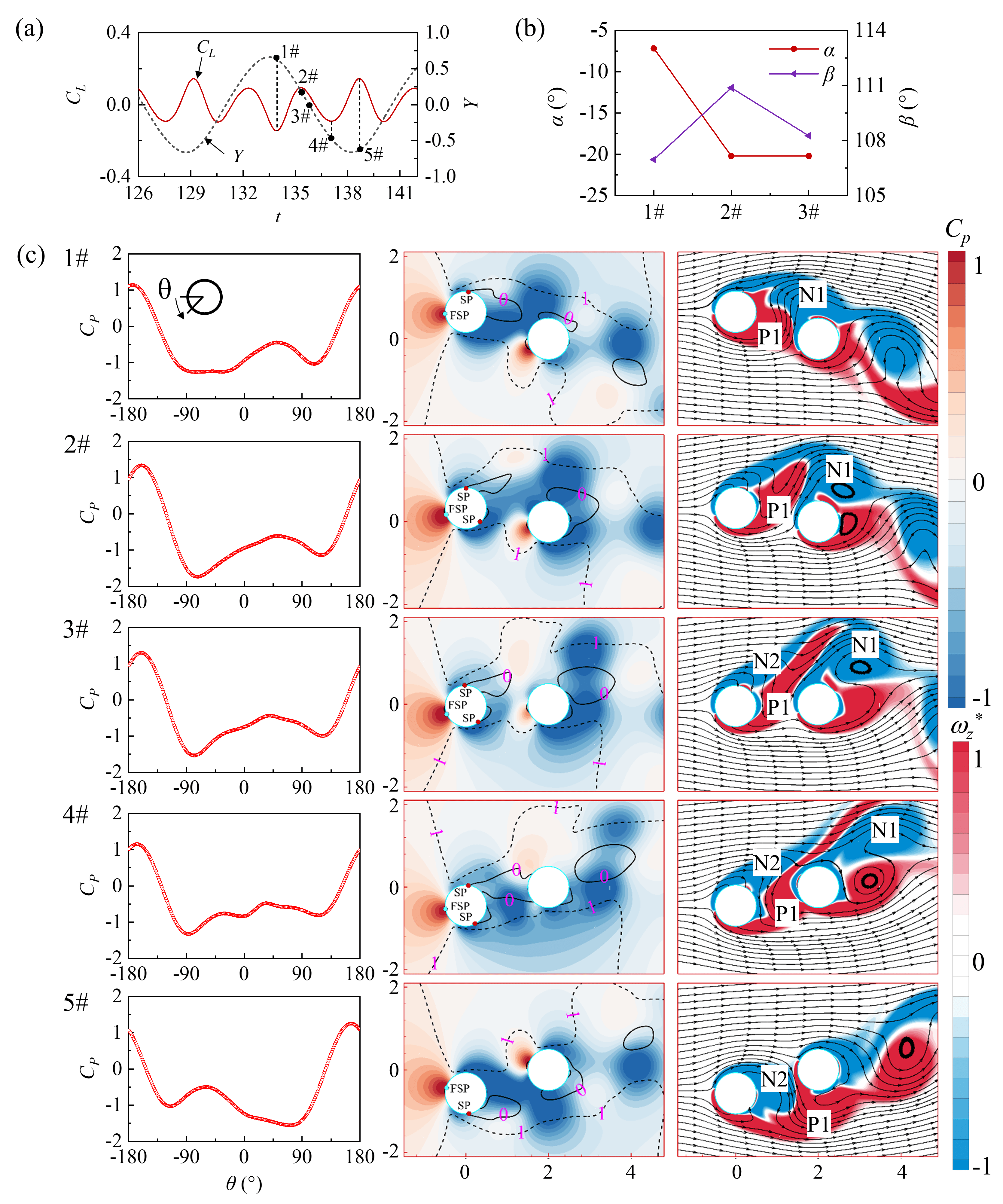}
    \caption{(a) lift coefficient and dimensionless amplitude, (b) variations of the $\alpha$ and $\beta$ in a half cycle of the displacement from 1\# to 3\#, (c) the surface pressure coefficient and the contour of the surface pressure coefficient and the local vorticity at different times when $Re = 160$.}
    \label{fig:fig23}
    \end{center}
\end{figure}

\hspace{0.5cm} Figures~\ref{fig:fig21}-\ref{fig:fig23} show the angle of the separation point of the boundary layer, the variation of the surface pressure coefficient ($C_P$), the contour of $C_P$ and vorticity at different instances of $Y$ and $C_L$ time histories for different $Re$. Fig.~\ref{fig:fig21}(a) shows the lift coefficient and dimensionless amplitude of the UC. In the $C_P$ contour, the solid and dotted lines represent the velocity $u^* = 0$ and $u^* = 1$, respectively, where $u^* = u/U_{in}$ is the normalized velocity. The separation point of the boundary layer can be determined from $u^* = 0$, while the distance between $u^* = 0$ and $u^* = 1$ indicates the thickness of the shear layer. Due to the symmetric motion of the UC, the angles of $\alpha$ and $\beta$ are only shown from 1\# to 2\# instances in Fig.~\ref{fig:fig21}(b). In Fig.~\ref{fig:fig21}(c), at 1\# instance, the separation point of the boundary layer is located on the upper surface of UC, and the angle $\beta$ between the separation point of the boundary layer and the front stagnation point is 118.7$^\circ$ (the definition of the angle is supplemented in Fig.~\ref{fig:fig1}). In the vorticity contours, the formation of the clockwise vortex N1 results in the maximum pressure difference between the upper and lower surfaces of the UC, in turn, the maximum upward lift coefficient at the 1\# instance. From the instances of 1\# to 2\#, the shear layer P1 moves with upward velocity and the intensity gradually increases, a small counterclockwise vortex is formed in the gap. Meanwhile, the clockwise vortex N1 begins to shed, and the supply of the boundary shear layer to the vortex N1 is weakened. The pressure on the upper surface of UC increases while the pressure on the lower surface decreases, causing the lift coefficient to gradually decrease to zero. From 2\# to 3\# instances, the counterclockwise vortex P1 causes a downward velocity in the gap, the pressure on the lower surface of the UC is lower than that on the upper surface, and the downward lift coefficient gradually increases. There is no harmonic interference effect at this $Re$.

\hspace{0.5cm} The harmonic components of $C_L$ emerge as the $Re$ value increases; see Fig.~\ref{fig:fig22}(a). It can be seen that the angle $\beta$ between the separation point of the upper surface boundary layer and the front stagnation point is smaller compared to that observed in $Re$ = 80; see Fig.~\ref{fig:fig22}(b). In Fig.~\ref{fig:fig22}(c), at 1\# instance, the clockwise vortex N1 induces an upward flow velocity in the gap and generates a positive lift force on the UC. At 2\# instance, the upper separation point of the boundary layer moves counterclockwise. The shear layer P1 is displaced to the rear side of the UC, and a small counterclockwise vertex is formed. The overall pressure level on the lower surface reduces, resulting in a decreased lift coefficient. From 2\# to 3\# instances, the counterclockwise vortex in the gap moves upward with the water flow, disrupting the connection between vortex N1 and the upper boundary layer of the UC. The separation point of the shear layer P1 moves downward, and the pressure level on the lower surface of UC increases, resulting in a slightly increased lift coefficient. At the 4\# instance, the connection between vortex N1 and UC weakens further, reducing the positive lift coefficient generated by N1 to 0. From the 4\# to 7\# instances, the growth of shear layer P1 generates a downward lift coefficient. At 6\# instance, the shear layer P1 causes a complete downward velocity in the gap. The connection between vortex N1 and UC is completely severed, the pressure on the upper surface of the UC increases to the maximum, and the downward lift coefficient generated by P1 reaches the maximum. At the 7\# instance, the enhancement of shear layer N2 leads to a decrease in the pressure on the upper surface and lift coefficient.

\hspace{0.5cm} At $Re$ = 160, the amplitude of the third harmonic frequency of the lift coefficient exceeds the amplitude of the fundamental vortex shedding frequency. As shown in Fig.~\ref{fig:fig23}(b), the angle between the front stagnation point and the boundary layer separation point is further reduced compared to that overserved in $Re$ = 120. At the 1\# instance in Fig.~\ref{fig:fig23}(c), the UC leave the positive maximum displacement position. The vortex N1 causes an upward flow velocity in the gap, and the positive shear layer P1 moves to the rear of the UC with the water flow. The time-history curves in Fig.~\ref{fig:fig23}(a) show that $C_L$ is minimum at this instance. From 1\# to 2\# instances, the shear layer P1 begins to cut off the connection between vortex N1 and the UC but is not destroyed. The separation point of the boundary layer on the lower surface moves downward, resulting in a decrease in the overall pressure on the lower surface of the UC. From 2\# to 4\# instances, the flow velocity in the gap turns downward, and the separation point of the lower boundary layer gradually moves downward. The pressure on the lower surface is still lower than that on the upper surface, and the lift coefficient gradually decreases with the variation of vortex P1. At the 4\# instances, the connection between vortex N1 and the UC is completely severed, causing an increase in the $C_L$ as the remaining shear layer N2 changes direction in the gap. Figures~\ref{fig:fig22} and \ref{fig:fig23} show that after the vortex is cut off by the shear layer in the gap, the growth of the remaining shear layer in the gap primarily controls the change in the lift coefficient. This remaining shear layer acts like a switch, and the variation in the lift coefficient depends on its position in the displacement curve. This behaviour differs from the third harmonic lift caused by the `2T' pattern.

\section{Conclusions}
In the present work, two-dimensional numerical simulations are carried out to study the influence of a fixed DC on the FIV of an elastically supported UC. The simulations are performed for a range of $Re$ from 50 to 170, and a range of $L/D$ from 1.25 to 3. The influence of $L/D$ and $Re$ on the FIV of UC is investigated in detail. The main conclusions are summarized as follows:

\begin{itemize}

    \item There is a significant impact of $L/D$ and $Re$ on the wake flow regime.Two distinct flow regimes, namely the steady flow, and alternating attachment, have been observed. The steady flow regime occurs in the low $Re$ regime. The $Re$ range for the steady flow regime decreases with increasing $L/D$. With the increase in $Re$ or $L/D$, the flow-field transitions from the steady flow regimes to an alternating attachment regime, which comprises the largest area in the $L/D-Re$ map.
    
    \item Various vortex shedding patterns are observed behind the tandem cylinders due to the hindrance of the DC on the vortex shedding of the UC. At $L/D$ = 1.25, two vortex shedding patterns, `2C’ and `aperiodic’, are observed. For $L/D$ is between 1.5 and 1.75, four different vortex shedding patterns are observed: `2S’, `2C’, `2P’, and `aperiodic’. At $L/D$ = 2, three different vortex shedding patterns, `2S’, `2C’ and `2P’, are observed. When $L/D$ is greater than 2, only the ‘2S’ and `2C’ patterns are observed.
    
    \item After the UC starts vibrating, the interference effect of the DC at a small $L/D$ slows down the increase rate of $\phi_{rms}$ between $C_L$ and $Y$. As a result, the lift of the UC continuously promotes the increment of the amplitude at a small $L/D$. When $L/D$ is between 1.25 and 1.75, as the velocity increases, the $\phi_{rms}$ remains less than $45^\circ$, and the corresponding vibration characteristic of the UC is characterized as interference galloping. The lock-in frequency of the UC increases with increasing $L/D$ but is smaller than the natural frequency of the UC. When $L/D \ge 2$, the vibration characteristics of UC show extended VIV. In this vibration regime, the lock-in interval of the cylinder is significantly larger than that of a single cylinder, and the amplitude of the lock-in region gradually decreases with increasing the $L/D$.
    
    \item The hindering effect of the DC on the vortex shedding of the UC results in variations in the shear layer in the gap, acting like a switch and altering the distribution of $C_P$ on the UC surface. When the $L/D$ is between 1.25 and 2, the third harmonic frequency of the $C_L$ of the UC increases with increasing $L/D$. Moreover, at $L/D$ = 2, the amplitude of the third harmonic frequency component will exceed that of the fundamental frequency component of $C_L$.
\end{itemize}

The findings of this study will allow us to better understand vibration reduction in tandem cylindrical structures and lead to the efficient design of flow-induced energy harvesters.

\section{Acknowledgements}
This work was supported by the National Natural Science Foundation of China (Grant No. 52277227), the Science and Technology Research \& Development Joint Foundation of Henan Province-Young Scientists (Grant No. 225200810099), and the Program for Science \& Technology Innovation Talents in Universities of Henan Province (Grant No. 23HASTIT010).

\section*{Data availability statement}
The data that support the findings of this study are available from the first author/corresponding author upon reasonable request.

\section*{Declaration of interests}
The authors report no conflict of interest.

\nocite{*}
\bibliography{manuscript}

\end{document}